\documentclass[aps,prb,twocolumn,showpacs,10pt]{revtex4-1}
\usepackage{graphicx}
\usepackage{amssymb}
\usepackage{amsmath,bm}
\usepackage{comment}
\usepackage{footmisc}
\usepackage{color}
\usepackage{epsfig}
\usepackage{graphicx,amsmath,amssymb,amsfonts,sidecap,color,hyperref}
\usepackage{subfigure}
\newcommand{\nn}{\nonumber\\}

\begin{document}

\title {From fractional boundary charges to quantized Hall conductance }

\author{Manisha Thakurathi}
\author{Jelena Klinovaja}
\author{Daniel Loss}
\affiliation{Department of Physics, University of Basel, Klingelbergstrasse 82, CH-4056 Basel, Switzerland}

\date{\today}

\begin{abstract}
We study the fractional boundary charges (FBCs) occurring in nanowires in the presence of periodically modulated chemical potentials and connect them to the FBCs occurring in a two-dimensional electron gas in the presence of a perpendicular magnetic field in the integer quantum Hall effect (QHE) regime. First, we show that in nanowires the FBCs take fractional values and change linearly as a function of phase offset of the modulated chemical potential. This linear slope takes quantized values determined by the period of the modulation and depends only on the number of the filled bands. Next, we establish a mapping from the one-dimensional system to the QHE setup, where we again focus on the properties of the FBCs.  By considering a cylinder topology with an external flux similar to the Laughlin construction, we find that the slope of the FBCs as function of flux is linear and assumes universal quantized values, also in the presence of arbitrary disorder. We establish that the quantized slopes give rise to the quantization of the Hall conductance. Importantly, the approach via FBCs  is valid for arbitrary flux values and disorder. The  slope of the FBCs plays the role of a topological invariant for clean and disordered QHE systems. Our predictions for the FBCs  can be tested experimentally in nanowires and in Corbino disk geometries in the integer QHE regime.
\end{abstract}

\maketitle

\section{Introduction}
Topological phases in condensed matter physics have gained considerable interest over the past decades, which was triggered by the experimental discovery of the integer as well as of the fractional quantum Hall  effect (QHE)[\onlinecite{Klitzing,Tsui,Laughlin,Girvin,Jain1,Jain2,TKNN,Mcdonald}]. Fractionalization of charges has been discussed in different topological systems and can emerge for various reasons. In the fractional QHE, strong electron-electron interactions are responsible for generating fractional excitations [\onlinecite{Read,Haldane,Halperin,English}]. However, fractional charges can occur also in non-interacting models as was first proposed in the Jackiw-Rebbi model [\onlinecite{Jackiw1},\onlinecite{Jackiw2}] and later in the Su-Schrieffer-Heeger model [\onlinecite{Ssh1,Ssh2,Goldstone}]. In these models, the fractional charge of $e/2$ is localized at domain walls [\onlinecite{Jackiw2},\onlinecite{Ssh2}]. Afterwards, such models were extended to describe also fractional charges localized at the boundaries [\onlinecite{Rice,Jackiw3,Kivelson}]. 
In contrast to fractional excitations in the fractional QHE, which were investigated in transport and shot noise experiments [\onlinecite{Steinberg,Inoue,Etienne,Mahalu}], the fractional boundary charges (FBCs) are far less explored experimentally, which is partially connected to the fact that the Jackiw-Rebbi and Su-Schrieffer-Heeger models are toy models. However, in recent years, there was a revival of  interest in FBCs with several models   being proposed that are realizable in condensed matter systems  [\onlinecite{Hughes,Goldman,Budich,Xu, Grusdt, DL1, Madsen,Poshakinskiy,DL2,DL3, Wakatsuki, Miert, Platero,Ryu,Marcel}].

\begin{figure}[t]
\centering
\includegraphics[width=0.75\columnwidth]{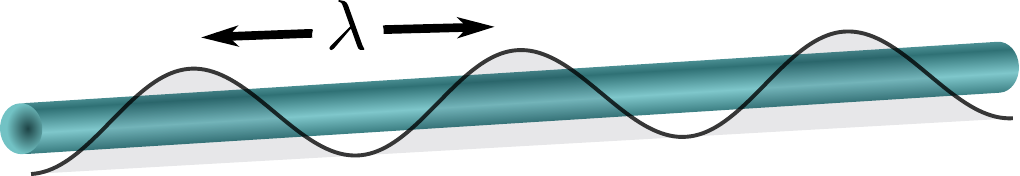}
\caption{Sketch of a one-dimensional NW (blue cylinder) in the presence of a chemical potential (black) which is periodically modulated, for example, by gates, with period $\lambda$.}
\label{fig01}
\end{figure}

In the present work, we first focus on the properties of FBCs in one-dimensional nanowires (NWs) with periodically modulated chemical potentials, see Fig. \ref{fig01}. Such a system is known to host in-gap bound states for a certain set of the offset phases $\alpha$, if the period of modulation $\lambda$ is tuned to  half of the Fermi wavelength, $\lambda=\pi/k_F$, where $k_F$ is the Fermi wavevector [\onlinecite{Gangadharaiah}]. However, as was shown subsequently, the FBCs in such setups do not not rely on the presence of such in-gap bound states and the FCBs are well-defined even if the bound states are absent [\onlinecite{Park}]. Remarkably, the FCBs in NWs turned out to be also very stable against moderate disorder [\onlinecite{Park}]. All these properties motivate us to study the FBCs in greater detail and, in particular, to generalize these findings to the regime in which the amplitude of the chemical potential modulation is comparable to the Fermi energy, thereby going beyond previous studies restricted to the perturbative regime [\onlinecite{Park}]. In addition, we consider regimes in which $\lambda$ is an integer multiple of half of the Fermi wavelength, $\lambda=\nu \pi/k_F$, with $\nu$ being a positive integer. Interestingly, also in this case, we find that there is a gap opening at the Fermi level. Moreover, this gap can host bound states if $\alpha$ is properly tuned. We also find that the FBCs are linear functions of $\alpha$ with the  slope $c_\nu = \nu e/2\pi$, being universal and quantized in units of $e/2\pi$. Again, this quantization is extremely robust against disorder, which suggests that this slope plays the role of a topological invariant for the system. 

In principle, the FBCs can be observed directly by using, for example, STM techniques to measure the charge at the boundaries of the NWs [\onlinecite{Park}]. In this way, one can also measure the linear dependence of the FBCs on the phase offset. However, we would like to connect the slope $c_\nu$ to other well-known quantized observables. For one-dimensional systems, the behaviour of the FBCs is connected to properties of  quantum charge pumps that transfer a quantized charge in each pumping cycle [\onlinecite{Thouless, Oded1, Marra,Wang,Niu1,Niu2,Oded2,Oded3}]. However, no such connection between FBCs and transport properties have been established yet in two-dimensional QHE setups. In this work, we attempt to fill this gap by connecting the quantized slope of the FBCs to quantized values of the Hall conductance in the integer QHE regime. To achieve this, we make use of the formal mapping between a 1D NW with periodic modulations and a 2D QHE system [\onlinecite{JK6}].  Such methods of dimensional extension or reduction were successfully employed to study properties of quasicrystals in different systems [\onlinecite{Oded1,Oded2,Oded3}].
If periodic boundary conditions are imposed along one of the two QHE boundaries, giving rise to a cylinder topology, the FBC can be controlled by flux insertion, thereby implementing the Laughlin setup [\onlinecite{Laughlin}]. Physical realizations of such a cylinder topology are given by Corbino disks in the QHE regime [\onlinecite{Jain1,Corbino,Syphers,Fontein,Dolgopolov,Zhu,Schmidt}].
Quite remarkably, the FBCs depend linearly on this flux and again with a slope $c_\nu$ that is universal and quantized like in the single NW case. We show that this slope quantization  is again very stable against  disorder in the whole sample (including the edges) as long as the bulk gaps are not closed. Finally, the quantized values of the slope  $c_\nu$ can be connected to the quantized values of the Hall conductance, $\nu e^2/h$. This connection clearly illustrates that all the occupied bulk states (via contributing to the FBCs) contribute to the Hall conductance and not just the edge states (which are responsible for the jump from one quantum Hall plateau to another).
In addition, the approach via FBCs shows that the Hall current changes continuously with an arbitrary change in the flux, 
unlike in the Laughlin argument where the Hall current is determined only for integer multiples of the flux quantum $\varphi_0=h/e$ [\onlinecite{Laughlin}]. 
Importantly, since our results are valid in the presence of disorder in the whole sample we can consider the universal quantized slope of the FBCs, $c_\nu$, as a  topological invariant in  integer QHE systems. The quantized slope can be accessed  by charge measurements, thus opening up alternative ways to study QHE systems  experimentally, beyond standard measurements via charge currents.

The outline of the paper is as follows. In Sec. \ref{sec2}, we introduce the model consisting of a single one-dimensional NW with periodically modulated chemical potential and calculate the FBCs for different values of the phase offset as well as for different number of filled bands. We identify characteristic features of the FBCs numerically and, in addition, provide analytical arguments to explain them. In Sec. \ref{sec3}, we map the aforementioned model to an integer QHE system consisting of an array of coupled NWs in the presence of magnetic field applied perpendicular to the NW plane. Next, we study the local particle density and the FBCs for the QHE system in the presence of an external flux  both in the absence (Sec. \ref{sec4}) and presence (Sec. \ref{sec5}) of disorder. In Sec.\ref{sec6}, we relate the quantized linear slopes of the FBCs to the quantization of the Hall conductance. Finally, in Sec. \ref{sec7}, we conclude with a summary and outlook.

\section{FBC in single Nanowire}
\label{sec2}

\subsection{Model} First, we consider a one-dimensional single-subband NW. Here and in what follows we neglect the spin degree of freedom and work with spinless electrons. The chemical potential is assumed to be periodically modulated, for example,  by external local gates, creating the charge density wave-type (CDW) modulation [\onlinecite{Deutschmann},\onlinecite{Algra}] with the amplitude $2V$ and period $\lambda$ along the entire length of the NW as depicted in Fig. \ref{fig01}. The tight-binding Hamiltonian of such a system has the following form
\begin{align}
&H_{1D}=-t_x\sum_{n=1}^{N-1} (\psi_{n+1}^\dagger \psi_{n}+\text{H.c.})\nn
&-\sum_{n=1}^{N}\left[ 2\,V\,\cos\left(2\,\pi\, n \frac{a_x}{\lambda} +\alpha \right)+\mu-2\,t_x \right]\, \psi_{n}^\dagger \psi_{n},
\label{Eq1}
\end{align}
where $\psi_n$ is an annihilation operator acting on an electron located at site $n$ of the NW of  length $l=(N-1)a_x$, with $N$ being the number of sites. The hopping amplitude $t_x$ and the lattice spacing $a_x$ determine the effective electron mass. The chemical potential $\mu$ is taken from the bottom of the band.  The phase offset $\alpha$ of the CDW, defined uniquely between $(-\pi,\pi]$, sets the value of the chemical potential at the left end of the NW (at site  $n=1$).

\begin{figure}[tb]
\centering
\includegraphics[width=0.75\columnwidth]{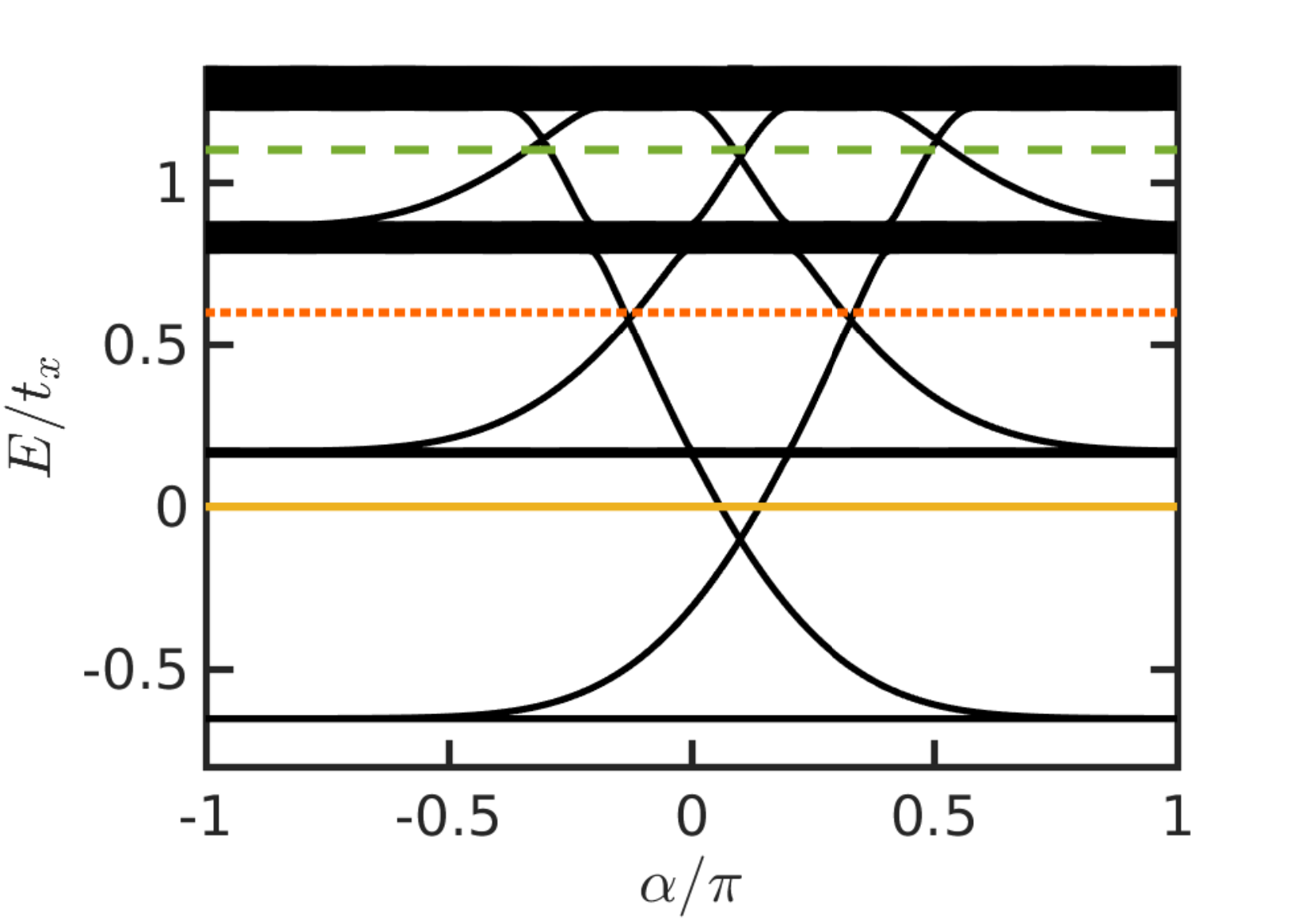}
\caption{ Energy spectrum of a  NW with CDW modulation of strength $V$ and period $\lambda$ as a function of phase offset $\alpha$.
Bulk gaps are opened by resonant scattering between the two Fermi points caused by the periodic modulation. If the chemical potential $\mu$ is tuned inside these gaps and the phase offset $\alpha$ is properly adjusted,  bound states  localized at the NW ends with energies inside the bulk gap emerge. The three colored horizontal lines correspond to the position of $\mu$ inside the first ($\mu/t_x=-0.1$, yellow solid line), second ($\mu/t_x=0.5$, orange dotted line), and third ($\mu/t_x=1$, green dashed line) bulk gap. The FBCs will be calculated at these values of $\mu$. The parameters are chosen as $N=600$, $V/t_x=0.6$, $\lambda/a_x=10$, and $\mu/t_x=-0.1$.}
\label{fig02}
\end{figure}

\subsection{ Energy spectrum} If  the CDW amplitude is small, $V \ll t_x$, we can  study the model analytically in the continuum regime [\onlinecite{Gangadharaiah},\onlinecite{JK2},\onlinecite{JK3}]. We begin with by linearizing the continuum model Hamiltonian close to the Fermi momenta $\pm k_F$, defined in terms of the chemical potential as $k_F a_x= \arccos(1-\mu/2t_x)$,  and by writing the fermion operators in terms of  slowly varying right and left movers denoted by $R(x)$ and  $L(x)$, respectively, as 
\begin{align}
\psi(x)= R(x)e^{i\,k_F\,x} +L(x)e^{-i\,k_F\,x}.
\end{align}
We neglect the fast oscillating terms and rewrite the kinetic part of the Hamiltonian  as
\begin{align}
H_{kin}=i \hbar\,v_F \int dx\, [L^\dagger(x)\,\partial_x\, L(x)-R^\dagger(x)\,\partial_x\, R(x)].
\end{align}
Here, $v_F$ is the Fermi velocity given by $\hbar v_F= 2\,t_x\,a_x^2\,k_F$. The CDW term has following form in the linearized model
\begin{align}
H_{CDW}= -V e^{-i\alpha}\int dx& \,[e^{2\,i\,x(\pi/\lambda-k_F)}+e^{-2\,i\,x(\pi/\lambda+k_F)}] \nn
&\times R^\dagger(x) L(x)+ \text{H.c.}
\label{CDW}
\end{align} 
Generally, such rapidly oscillating terms average out to zero unless the resonance condition $k_F= \nu \, \pi/\lambda$,  with $\nu$ being an integer, is satisfied. We first analyze the special case $\nu=1$ and then consider general $\nu$.
In this case, $H_{CDW}$ couples  right and left movers at the Fermi level, and as a result a gap of  size $\Delta_g^{(1)}=V$ opens in the spectrum [\onlinecite{Gangadharaiah}], see Fig. \ref{fig02}. 

In the basis $\tilde\psi=(R,L)$, the total linearized Hamiltonian in the resonance case ($k_F=\pi/\lambda$) has the form $H_{1D}=\int dx \,\psi^\dagger \mathcal{H} \,\psi$  with Hamiltonian density $\mathcal{H}=\hbar\,v_F \hat k\,\sigma_z- V\cos(\alpha) \sigma_x+ V\sin(\alpha)\sigma_y$, where $\hat k= -i\, \partial_x$ is the momentum operator with eigenvalue $k$. The bulk spectrum is given by  $E_{\pm}= \pm [(\hbar v_F k)^2 +V^2]^{1/2}$. For an infinitely long NW, no states reside inside the bulk gap as the spectrum is fully gapped for all values of $k$.  To explore the possibility of bound states in the gap [\onlinecite{JK4}], we consider a finite NW of length $l$ with the condition that $l \gg \xi$ [\onlinecite{JK5}], where $\xi$ is the localization length of the bound state. Next, we impose  vanishing boundary condition at the left (right) end of the NW, $x=0$ ($x=l$), such that $R(0)+L(0)=0$ [$R(l)+L(l)e^{-2ik_Fl}=0$]. The spectrum of the bound state localized at the left (right) boundary of the NW depends on the phase offset $\alpha$ and is given by $\epsilon(\alpha)=V\,\cos(\alpha)$ [$\epsilon(\alpha)=V\,\cos(\alpha+2k_Fl)$] under the constraint  $ \sin(\alpha)<0$ [$ \sin(\alpha+2k_Fl)<0$]. If the latter constraints are not satisfied, there is no bound state. The corresponding wavefunctions have the form $\tilde\phi \sim \sin(k_F\, x)\, \text{exp}(-x/\xi)$ [$\tilde\phi \sim \sin(k_F\, (x-l))\, \text{exp}[(x-l)/\xi]$]  with the localization lengths defined as $\xi=-\hbar v_F/[V\,\sin(\alpha)]$ [$\xi=-\hbar v_F/[V\,\sin(\alpha+2k_Fl)]$].

Next, we can generalize these result to arbitrary positive integer $\nu$ where  the condition $k_F= \nu \, \pi/\lambda$ also allows for resonant scattering between left and right movers in higher orders of perturbation theory [\onlinecite{M1},\onlinecite{M2}]. In this case, the gap is opened in the $\nu^{th}$-order of perturbation expansion and is of order of $\Delta_g^{(\nu)} \approx V^{\nu}/E_0^{\nu-1}$, where $E_0$ is the characteristic energy which depends on the chemical potential of the system.  We refer to Appendix {\ref{ap1}} for further details. The gap $\Delta_g^{(\nu)}$ is reduced by a factor $(V/E_0)^{\nu-1}$ in comparison with the direct gap $\Delta_g^{(1)}$, which implies that as the value of $\nu$ increases the gap decreases.  The spectrum of the bound states can also be calculated in a similar way as done before for $k_F=\pi/\lambda$. However, we note here that one can directly recalculate the energy spectrum by rescaling $\alpha\rightarrow \nu\alpha$ and also $V \rightarrow \Delta_g^{(\nu)}$ [see  Appendix {\ref{ap1}} for more details].  As an important consequence, in this perturbative regime,  the spectrum of bound states at one given NW end satisfies $\epsilon (\alpha)=\epsilon(\alpha+2\pi p/\nu)$ for $p=0,...,\nu -1$. This feature ensures that as one changes $\alpha$ from $-\pi$ to $\pi$, there will be $\nu$ bound states, localized at each NW end, at any given energy inside the $\nu$th bulk gap, see Fig. \ref{fig02}.

\begin{figure}[t]
\centering
\includegraphics[width=0.65\columnwidth]{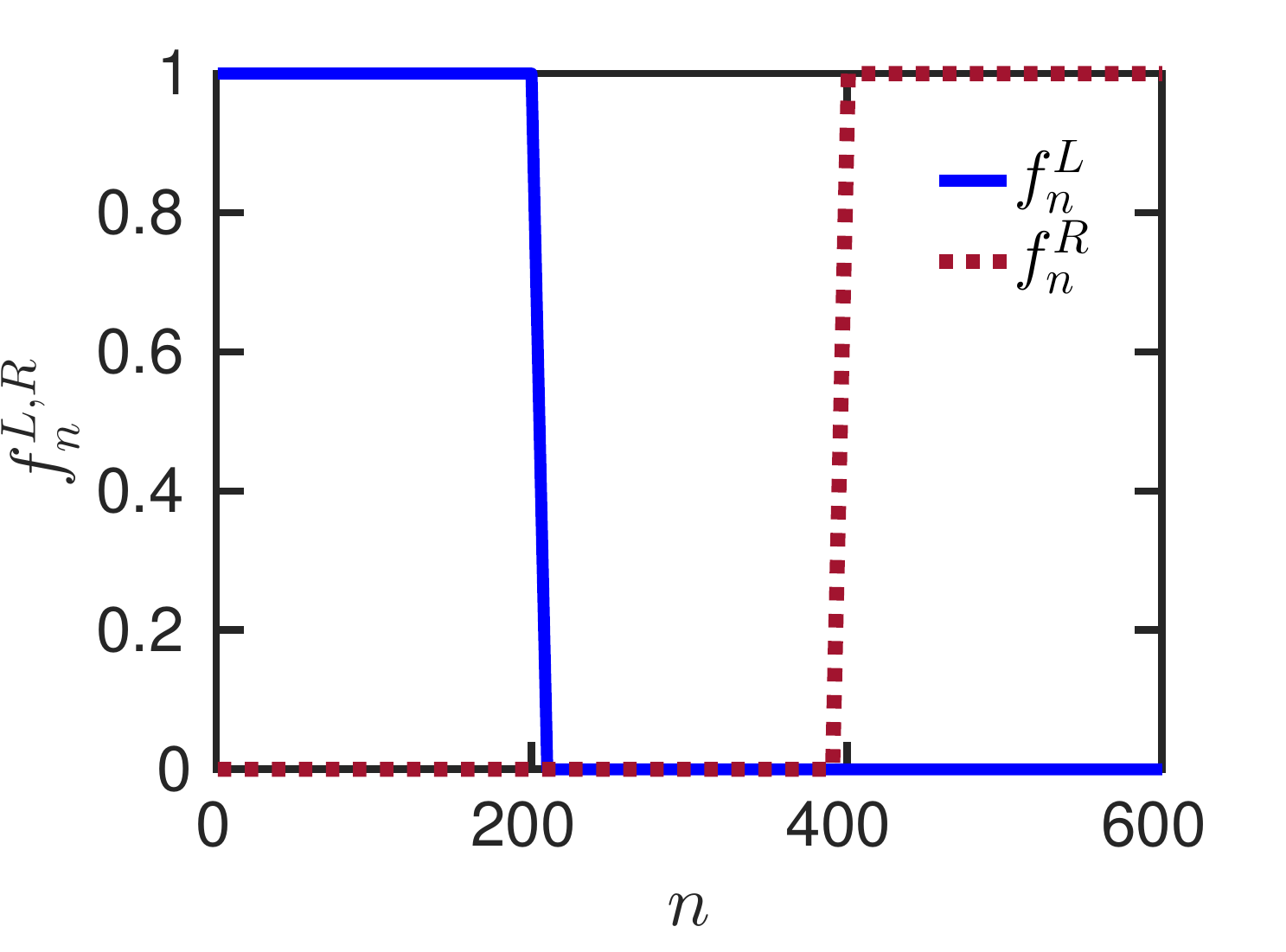}
\caption{The profile function $f^L_n$ ($f^R_n$) is defined to capture features of the FBC at the left (right) NW end as a function of lattice site $n$. The used parameters are $N=600$, $n_1=200$ and $n_2=10$.}
\label{fig03}
\end{figure}

As the amplitude of the CDW grows, $V \simeq t_x$, the CDW cannot be treated perturbatively anymore. The size of the bulk gaps gets larger compared to the perturbative regime up to the point at which the energy bands get flat, see Fig. \ref{fig02}.   However, as the bulk gap never closes upon increasing $V$, one can conclude that the bound states are still present in the spectrum and, moreover, their number inside a given gap is also not changing. When the chemical potential lies inside the lowest gap, the bound states obtained above were discussed before in different contexts in Refs. [\onlinecite{Tamm,Shockley,Gangadharaiah}]. 
Here, we have shown in addition that the number of bound states increases as one tunes the chemical potential inside the bulk gaps opened at higher energies. Furthermore, we also note that the commensurability relation between $\lambda$ and $a_x$ does not play any role in our setup, which is also confirmed by the analytical solutions obtained in continuum limit.

\begin{figure*}[t]
\begin{center} \begin{tabular}{ccc}
\epsfig{figure=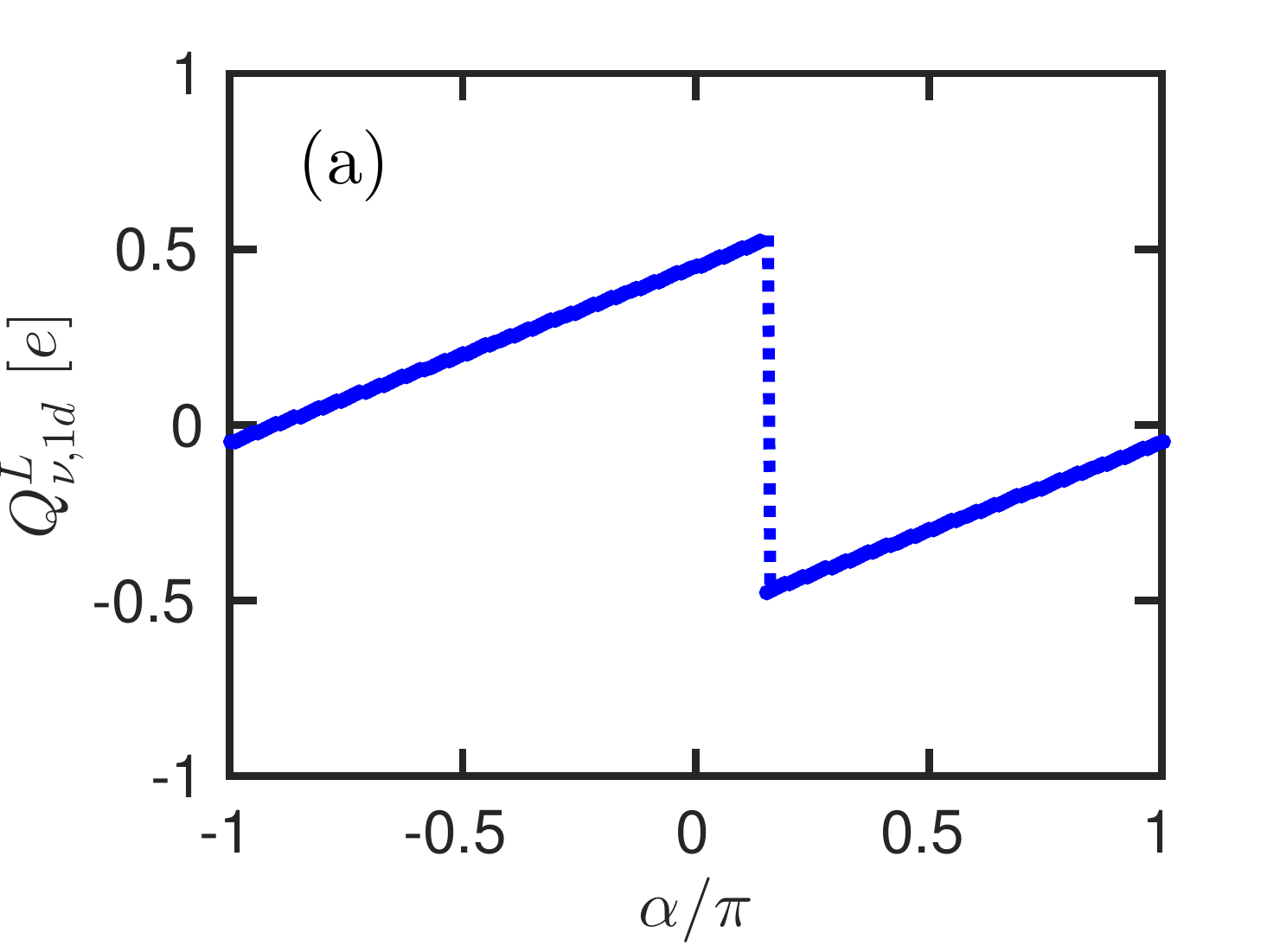,width=2.3in,height=1.8in,clip=true} &\hspace*{-0.cm}
\epsfig{figure=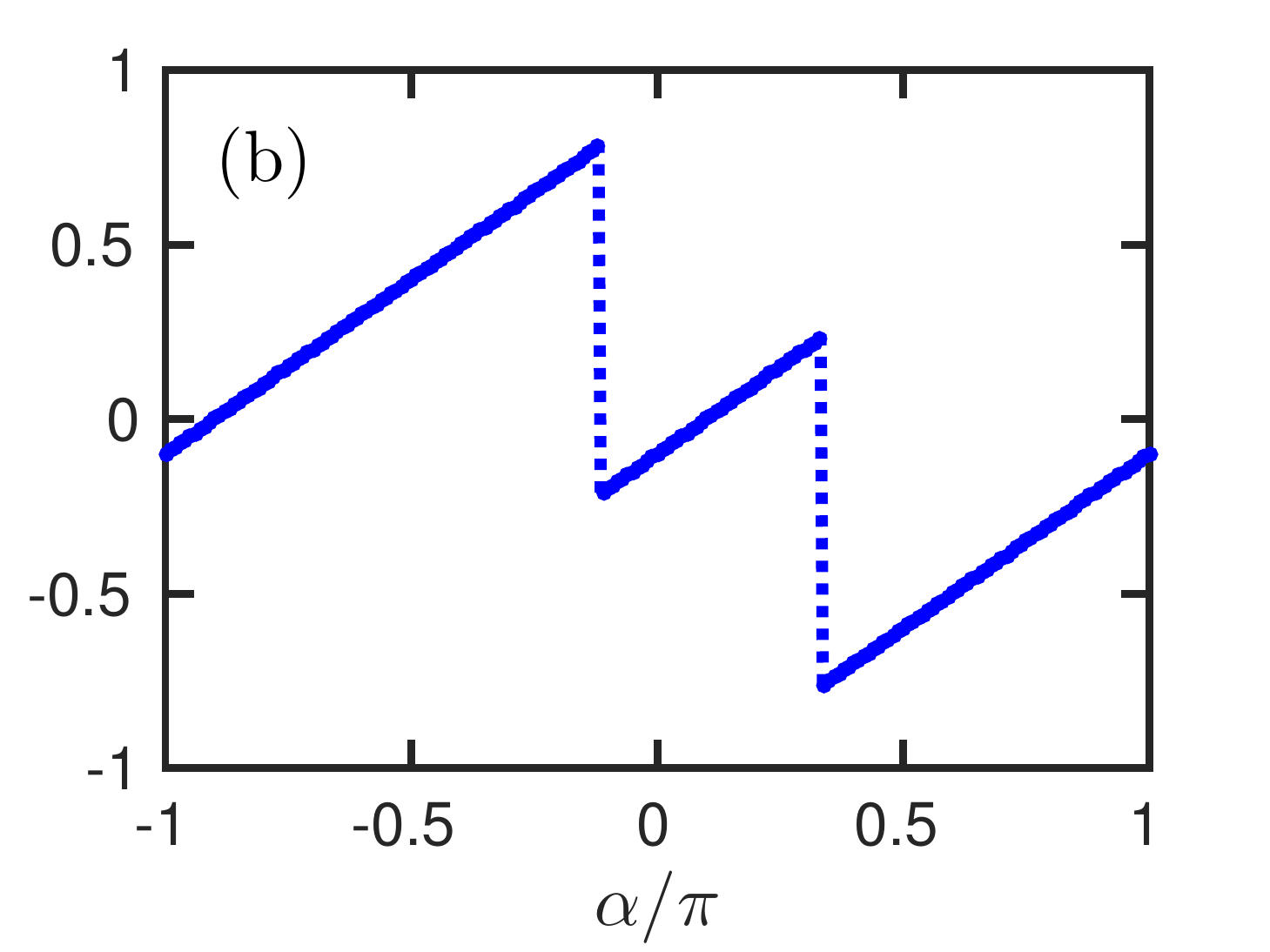,width=2.2in,height=1.8in,clip=true} &\hspace*{-0.cm}
\epsfig{figure=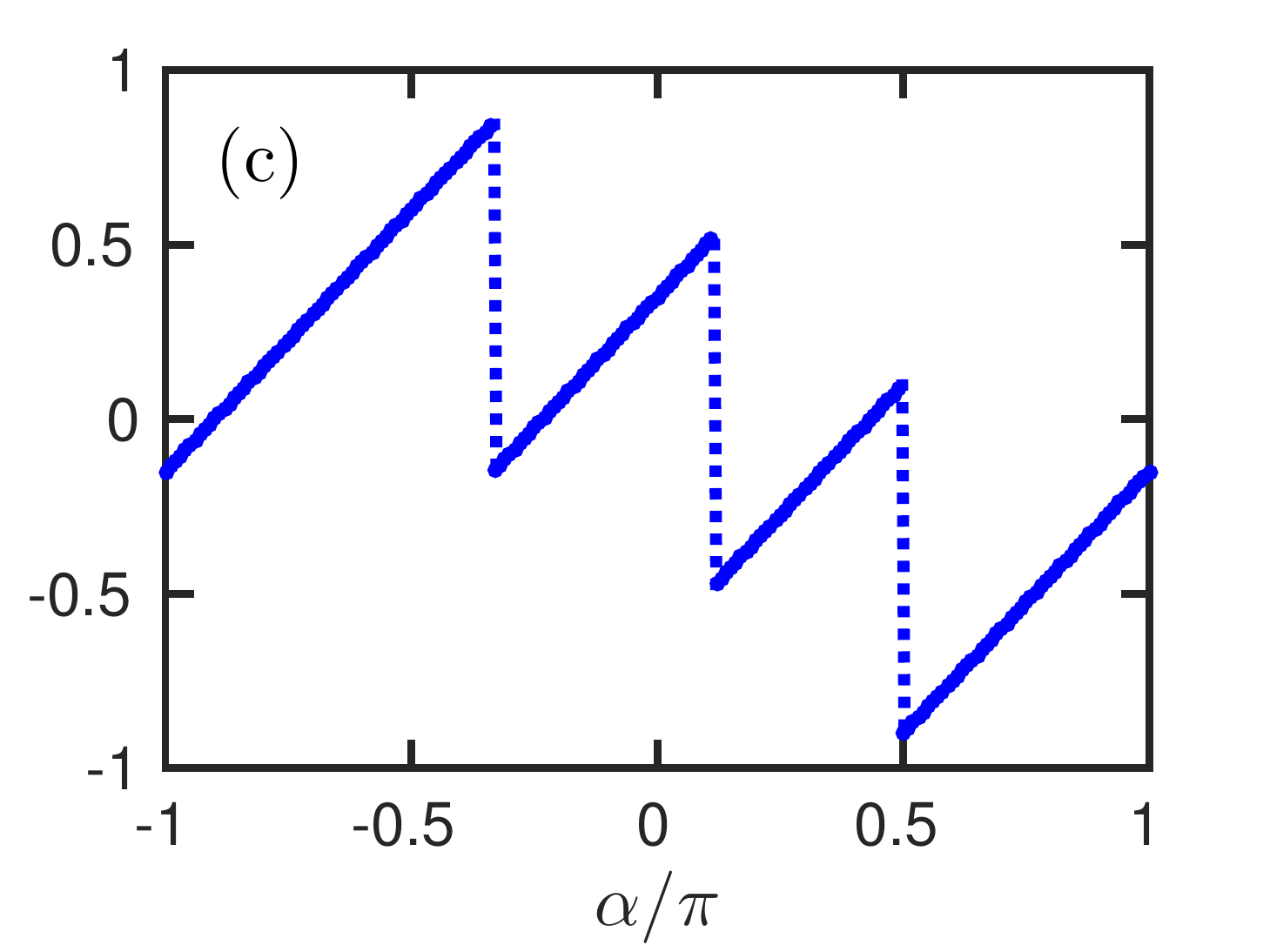,width=2.2in,height=1.8in,clip=true} \\
\epsfig{figure=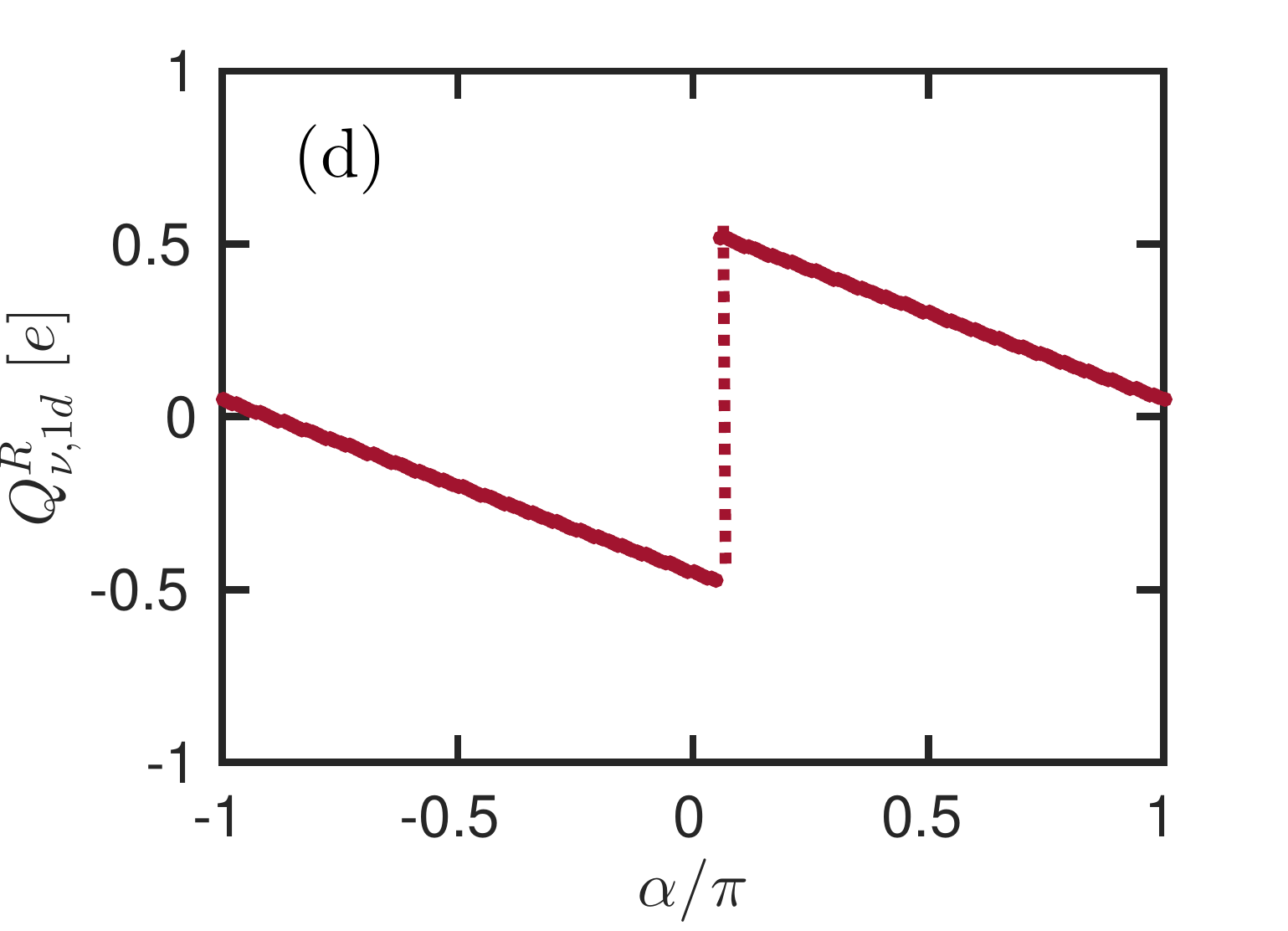,width=2.3in,height=1.8in,clip=true} &\hspace*{-0.cm}
\epsfig{figure=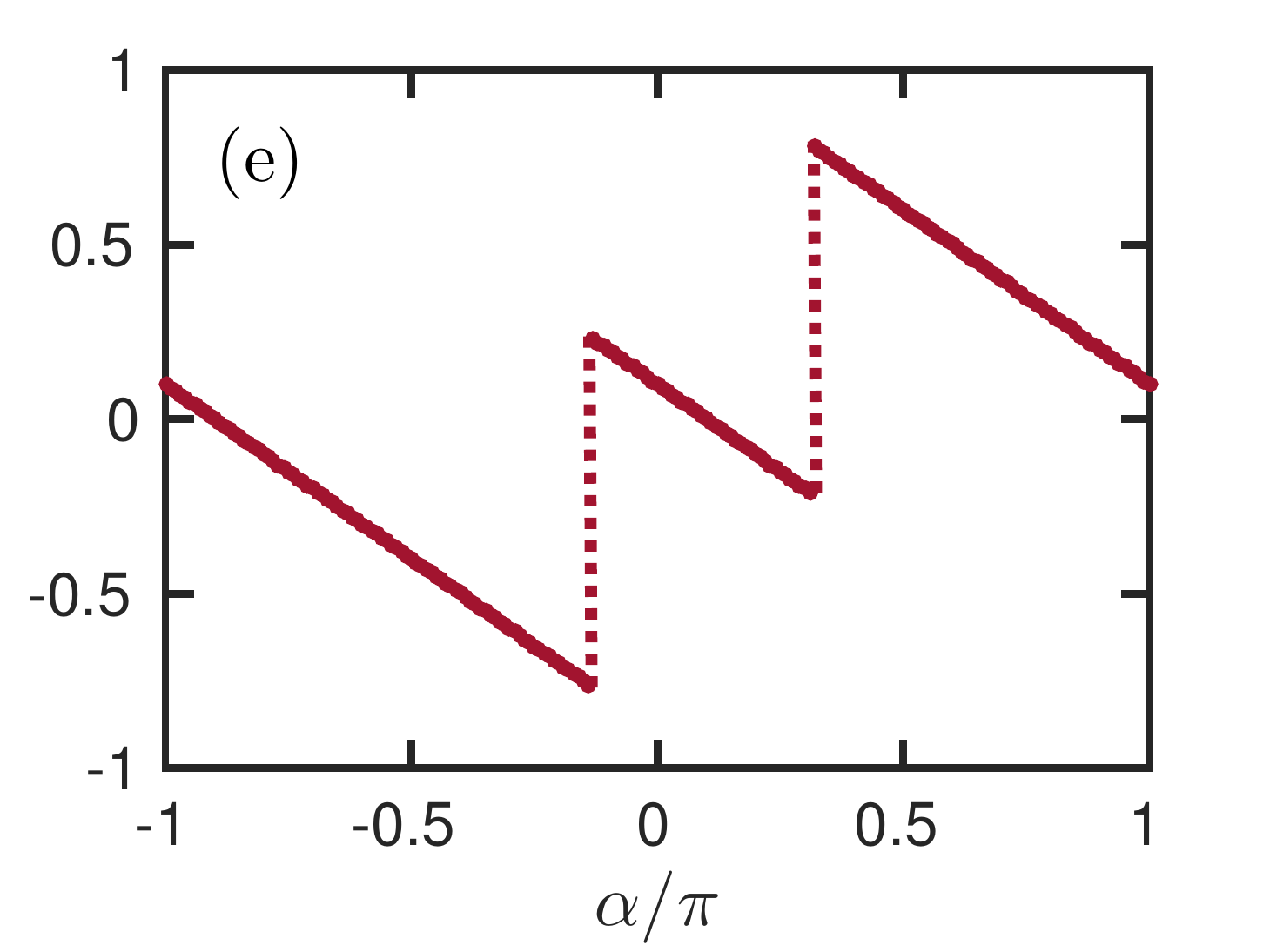,width=2.2in,height=1.8in,clip=true} &\hspace*{-0.cm}
\epsfig{figure=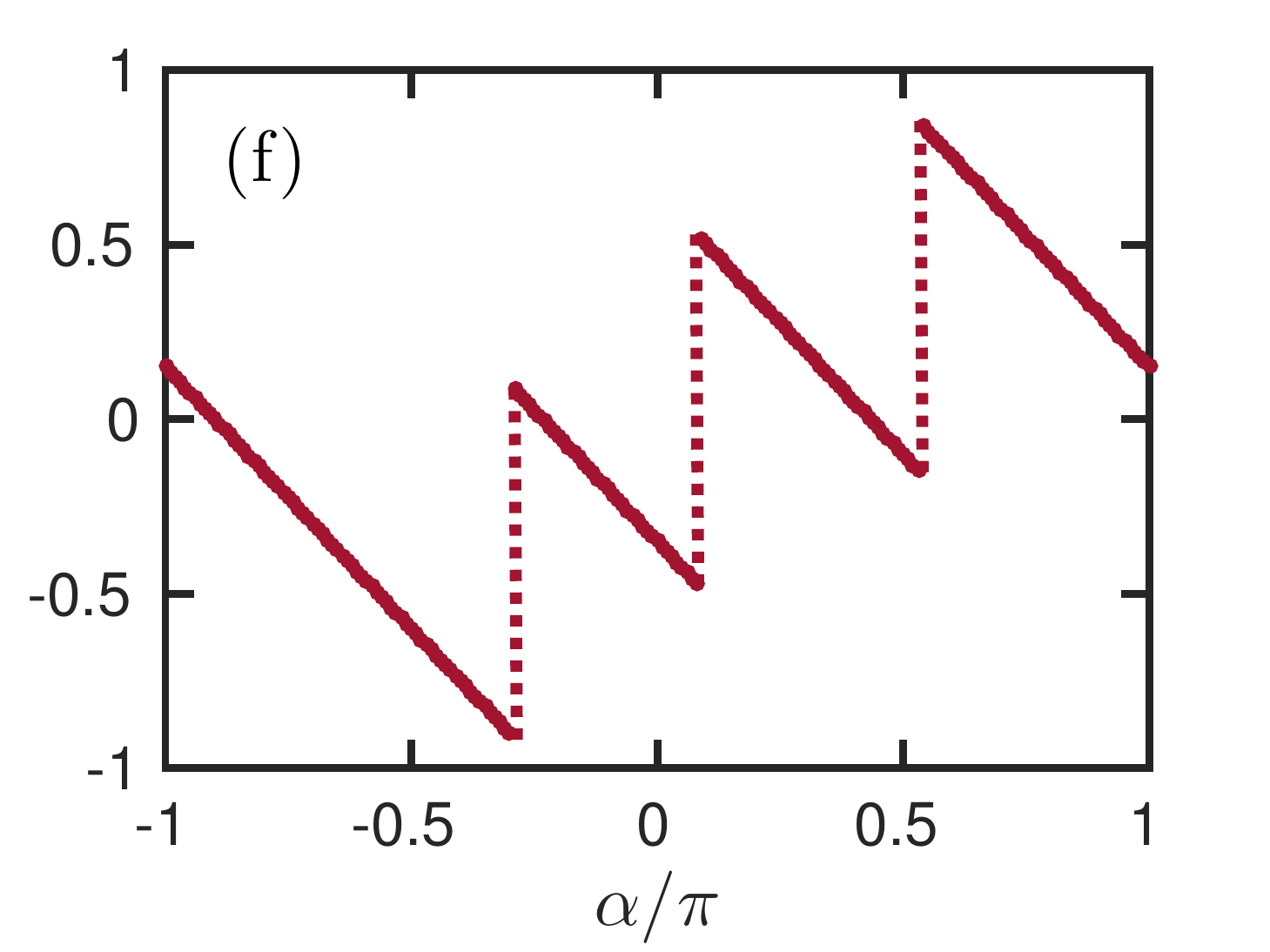,width=2.2in,height=1.8in,clip=true} 
\end{tabular} \end{center}
\caption{The FBCs $Q_{\nu,1d}^L$ [blue, (a)-(c)] and $Q_{\nu,1d}^R$ [red, (d)-(f)] measured in the units of electron charge $e$ as a function of the phase offset $\alpha$ obtained numerically for the CDW-modulated NW. The number of filled bands is controlled by the chemical potential that is tuned inside the first [(a),(d)], second [(b),(e)], or third [(c),(f)] bulk gap, see Fig. \ref{fig02}. The FBCs vary linearly as a function of $\alpha$ and the absolute value of the slope is quantized for (a,d), (b,e), and (c,f) as $1/2\pi$, $2/2\pi$, and $3/2\pi$. The sign of the slope is opposite at two boundaries. The FBCs jump by $\pm e$ at the bound states cross the chemical potential.
The parameters are chosen to be the same as in Figs. \ref{fig02} and \ref{fig03}.
}
\label{fig04}
\end{figure*}

\subsection{Fractional boundary charge} Next we turn to the FBC in a single NW with CDW modulation. To begin with,  we define the FBC at each of two NW ends as [\onlinecite{Park}]
\begin{align}
 Q_{\nu,1d}^s= \sum_{n=1}^{N} f^s_n (e \, \langle \psi_n^\dagger \psi_n \rangle-\bar \rho_\nu).
\label{Eq2}
\end{align} 
where we have subtracted from the expectation value of the charge density in the the ground state at site $n$, $ e\,\langle \psi_n^\dagger \psi_n \rangle$,  the average bulk charge per site, $\bar \rho_\nu$. Here, $e$ is the electron charge. If the chemical potential is tuned inside the $\nu$th bulk gap, we have $\bar \rho_\nu=e\,\nu\, a_x/\lambda$. 
To capture the FBCs at the left ($s=L$) and right ($s=R$) NW boundaries separately, we introduced  a profile function $f^s$, which has spatial support only at one of the two NW boundaries, see Fig. \ref{fig03}. Without loss of generality, we work with the following profile function defined by two cut-offs $n_1,n_2>0$ (for sharp transition, $n_2=1$),
\begin{align} \label{frl}
f_n^L=&\Theta(n_1+n_2-n)\nn
&-\frac{n-n_1}{n_2} [\Theta(n-n_1)-\Theta(n-n_1-n_2)],
\end{align}
where $\Theta(x)$ is the heaviside step function. The profile function at the right, $f_n^R$, is mirror symmetric to the function at the left  given by $f_{n}^L=f_{N+1-n}^R$, where $1\leq n\leq N$. A well defined FBC should be independent of the form of the profile function and of the precise choice of the two cut-offs as long as $n_1 a \gg \xi$ and $n_2 a \gg \lambda$. Note that {\it all} states filled up to the Fermi level contribute to the FBC including a possible bound state.

After defining the FBCs, we calculate it numerically for the left and right NW boundaries and for different positions of the chemical potential $\mu$ inside the $\nu$th gaps, see Fig. \ref{fig04}. We observe the following four salient features: (1) The FBCs change linearly as a function of the phase offset $\alpha$, which allows us to define the slope of the linear function describing this dependence. (2) The slope is given strictly by the universal value $\pm e\nu/2\pi$. Thus, the slope is quantized and depends only on the number of filled bands, or in other words, on the band gap inside which the chemical potential is tuned. The position of the chemical potential inside the band gap does not affect the slope. The sign of the slope defined for the right and left FBCs are opposite. (3) The FBCs change continuously and can take positive and negative values. These values are usually bounded between $-e$ and $e$. (4) The FBCs jump by the amount $\pm e$ as the energy of the bound states localized at the corresponding NW end flips its sign, as one changes $\alpha$. Indeed, if the bound state energy is negative (positive), the corresponding state is filled (empty), and, thus, it contributes (does not contribute) with charge $e$ to the FBCs. Consequentially, as the bound state crosses the  chemical potential, the FBCs should change by $\pm e$. The position of such jumps depends on the precise position of the chemical potential inside the $\nu$th gap. In contrast to that, the number of jumps is determined by the number of bound states and is quantized and given by $\nu$.

\subsection{ Linear dependence of FBCs on phase offset}
\label{ana}
In this subsection we discuss the functional dependence of FBCs on the phase offset $\alpha$ and provide analytical arguments to support the linear dependence between these two quantities established numerically in the previous subsection. For this we need to generalize the approach given in 
Ref. [\onlinecite{Park}] for
$\nu=1$ to arbitrary integer values of $\nu$.
For simplicity, we carry out the proof in the tight-binding model description, where we also assume that $\lambda$ and $a_x$ are commensurable. However, this is not a crucial requirement and this constraint can be loosened if one switches to the continuum description. We define the total charge $Q_{1d}$ of the NW decomposed into three parts [\onlinecite{Park}],
\begin{align}
Q_{1d}=Q_{\nu,1d}^b +Q_{\nu,1d}^L+Q_{\nu,1d}^R,
\label{Eq5}
\end{align}
where $Q_{\nu,1d}^b=N\bar \rho_\nu = N e\, \nu\,a_x/\lambda$ is the charge of the constant (uniform) bulk background and  $Q_{\nu,1d}^{L/R}$ the FBCs at the left/right boundary of the NW.
In what follows the chemical potential $\mu$ is assumed to be inside the $\nu$th bulk gap, such that $Q_{1d}$ is an integer multiple of $e$. Below we study the change in the FBCs $Q_{\nu,1d}^{L,R}$ upon changing the system size $N$ and the phase offset $\alpha$. First, we note that in long NWs $ Q_{\nu,1d}^R$ does not change if one extends the NW by one full period of the CDW, {\it i.e.}, by changing the size from $N$ to $N+\lambda$. Thus, $Q_{\nu,1d}^R$ must be a function of $\delta= N\, {\rm mod} (\lambda/a)$.  Let us now consider the following steps.

(1) We extend the NW at the right end by one site such that the number of sites increase to $N+1$.  Therefore, the bulk charge $Q_{\nu,1d}^b$ increases by $\nu\, e\,a_x/\lambda$. As $Q_{1d}$ can take only integer values, this change should be compensated by $Q_{\nu,1d}^b$, and, thus, the FBCs $Q_{\nu,1d}^{L,R}$ have to decrease by $\nu\, e\,a_x/\lambda$. However, the FBC at the left NW end $Q_{\nu,1d}^L$ should remain unaffected by manipulations on the right NW end, thus, 
\begin{align}
 Q_{\nu,1d}^R(N+1)- Q_{\nu,1d}^R(N)= -\nu\, e\,a_x/\lambda.
\label{Eq6}
\end{align} 
Here, the change in the FBC is defined up to $\pm e$.

(2) We note that if one readjusts $\alpha$, one can compensate for the shift of the right boundary by one site and keep $ Q_{\nu,1d}^R$ unchanged. This would require to change $\alpha$ as  $\alpha \rightarrow \alpha - 2 \,\pi\,a_x/\lambda$. 
Thus, $Q_{\nu,1d}^R$ is not a function of two independent parameters $\alpha$ and $\delta$ but only of their combination, {\it i.e.},
\begin{align}
Q_{\nu,1d}^R(\delta, \alpha)= Q_{\nu,1d}^R\left(\frac{a_x\,\delta}{\lambda}+\frac{\alpha}{2\pi}\right).
\label{Eq7}
\end{align}
From Eq. (\ref{Eq6}) we conclude that $Q_{\nu,1d}^R$ is a linear function of $\delta$. Hence, it follows from Eq. (\ref{Eq7}) that the FBC $Q_{\nu,1d}^R$ is also a linear function of $\alpha$,
\begin{equation}
Q_{\nu,1d}^R (\alpha)=- c_\nu\ \alpha + C^R,
\label{QR}
\end{equation}
where the slope is determined by $c_\nu=e\, \nu/2\pi$, in full agreement with our  numerical findings, see Fig. \ref{fig04}. The piecewise constant function $C^R$ will be neglected in what follows. We just note that $C^R$ jumps by $e$ as one of the bound states crosses the chemical potential upon changing $\alpha$. The number of the bound states at each end is given by $\nu$. Thus, the total change of $C^R$ as $\alpha$ is changed continuously by $2\pi$ is $e \nu$. 
This ensures the periodicity of the FBC, $Q_{\nu,1d}^R(\alpha)=Q_{\nu,1d}^R(\alpha+2\pi)$.

As one changes $\alpha$, the bulk contribution $Q_{\nu,1d}^b$ to the total charge stays constant. Thus, the sum of the two FBCs, $Q_{\nu,1d}^R+Q_{\nu,1d}^L$, must also remain unchanged unless there is a bound state crossing the chemical potential. This means that the left FBC $Q_{\nu,1d}^L$ is also a linear function of $\alpha$ with the same absolute value of the slope $c_\nu$. However, the sign of the slope is opposite such that $Q_{\nu,1d}^L$ ($Q_{\nu,1d}^R$) increases (decreases) as $\alpha$ is increased. This is again in full agreement with the numerical results,  see Fig. \ref{fig04}.

\begin{figure}[b]
\centering
\includegraphics[width=0.9\columnwidth]{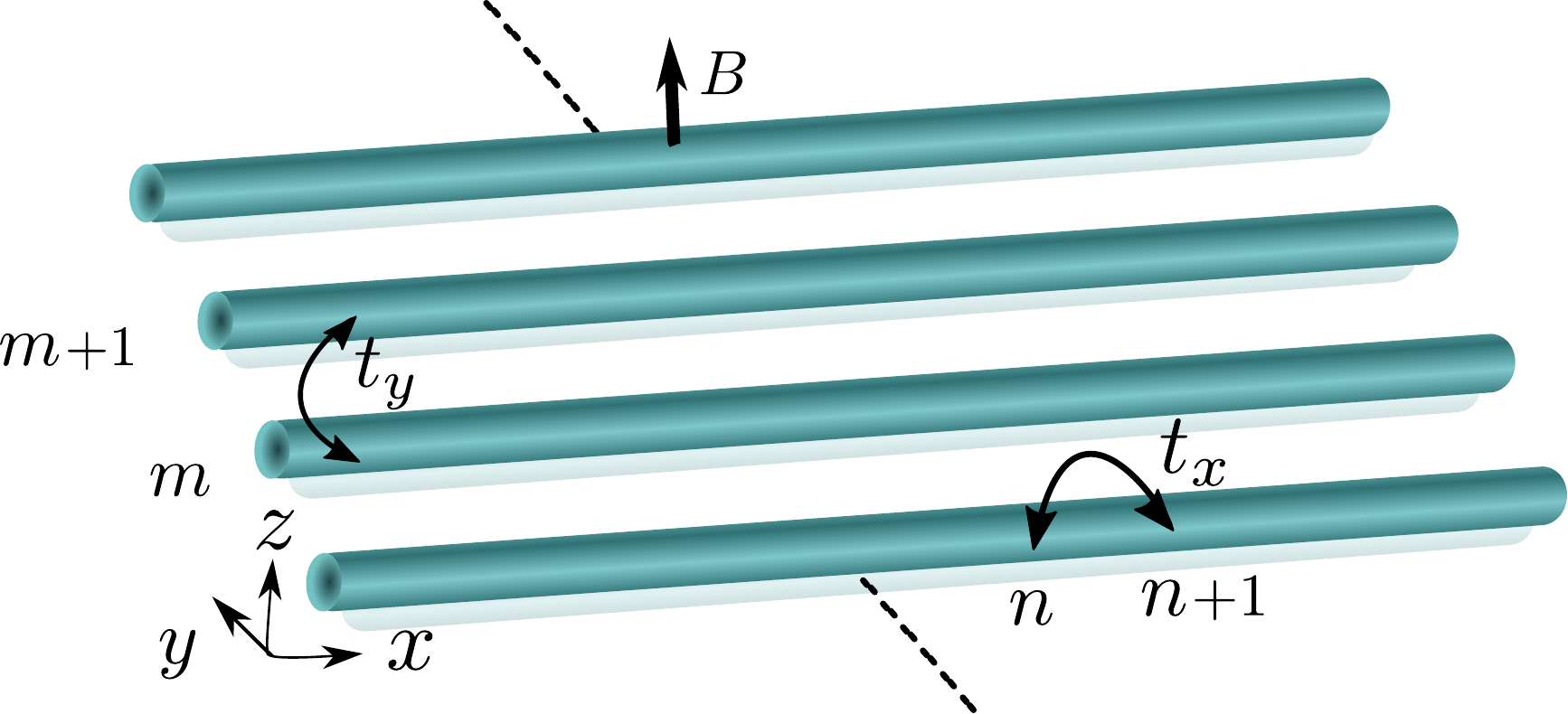}
\caption{An array of one-dimensional tunnel-coupled NWs (blue cylinders) is placed in the $xy$ plane.
The external magnetic field ${\bf B}$ is applied along the $z$ direction. The electron tunnels with an amplitude $t_x$ ($t_y$) within the NW (between two neighbouring NWs). The array models an integer QHE  for appropriate values of the $B$ field.}
\label{fig05}
\end{figure}

\section{Mapping from NW to QHE system}
\label{sec3}

We extend now our considerations to 2D systems in the QHE regime. First we consider the clean case and subsequently add disorder.
The setup considered in previous Sec. \ref{sec2}, which consists of a single NW with periodically modulated chemical potential, can be mapped to a system of tunnel-coupled NWs in a uniform magnetic field as follows [\onlinecite{Yakovenko,JK6,JK1,Oded1,Kane1}]. We consider a finite array of $M$ tunnel-coupled NWs, in the presence of a magnetic field which is applied perpendicular to the plane of the NWs, {\it i.e.}, along the $z$ direction, as shown in Fig. \ref{fig05}.  We work in the Landau gauge and choose the corresponding vector potential to be along $y$ direction, in Cartesian coordinates, ${\bf A}=B\,x\,\hat{y}$. Therefore, the Peierls phase, which the electron accumulates as it tunnels between NWs, is given by   
 $\phi(x) = (e/\hbar)\int {\bf A}\cdot {\rm d}{\bf  l}= (eB\,a_y/\hbar)\,x=(2\,\pi/\lambda)\, x$, where we have introduced $\lambda=h/eB\,a_y$ and used ${\rm d}{\bf  l}=dx\, \hat x+ dy\, \hat y+ dz\, \hat z$. In the discretized model of the NW consisting of $N$ sites, we have $x=n\,a_x$, with $1\le n\le N$ being an integer. Here, $a_x$ and $a_y$ are the lattice spacings along $x$ and $y$ direction, respectively.

\begin{figure}[b]
\centering
\includegraphics[width=0.75\columnwidth]{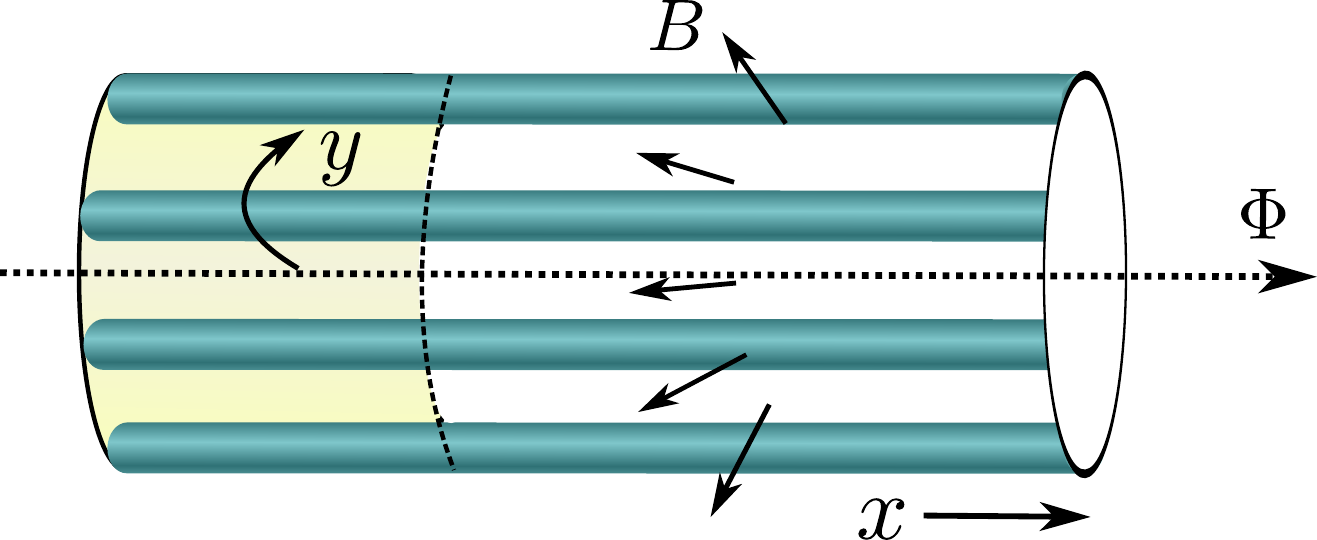}
\caption{Sketch of a periodic array of tunnel coupled NWs (blue) with the topology of a cylinder oriented along $x$ direction. The magnetic field $B$ points normal to the cylinder surface and is responsible for bringing the setup into the QHE regime. An external flux $\Phi$ is applied along the cylinder axis and, when changing in time, induces an electromotive force $\mathcal{E}_y$ in azimuthal direction $y$. For the calculation of the FBC and the Hall conductance, we focus on the patch (yellow) of area $\cal{A}$ localized at the left boundary of the system.}
\label{fig06}
\end{figure}
The corresponding tight-binding Hamiltonian for this NW array is given by
\begin{align}
H=\Big[&-t_x\sum_{(n,m)=(1,1)}^{(N-1,M)} \psi_{n+1,m}^\dagger \psi_{n,m}\nn&-t_y \sum_{(n,m)=(1,1)}^{(N,M-1)} e^{i \,2\,\pi\,n\,a_x/\lambda}\psi_{n,m+1}^\dagger  \psi_{n,m} \Big]+\text{H.c.}\nn
&-(\mu-2\,t_x) \sum_{(n,m)=(1,1)}^{(N,M)}\psi_{n,m}^\dagger \psi_{n,m},
\label{Eq3}
\end{align}
where $\mu$ is the chemical potential and $t_x$ and $t_y$ are the hopping amplitudes inside each of the NW and in between  two neighboring NWs, respectively. The annihilation operator $\psi_{n,m}$ acts on an electron located at site $n$ of the $m$th NW.

Next, we impose periodic boundary conditions along the $y$ direction and introduce tunneling $t_y$ also between 
the first and $M$th NWs, see Fig. \ref{fig06}. Thus, the momentum $k_y$ defined along the $y$ direction is a good quantum number and takes quantized values ranging from $-\pi/a_y$ to $\pi/a_y$ in steps of $ 2\, \pi/(M\,a_y)$. Applying the Fourier transformation, $\psi_{n,m}= (1/ \sqrt{M})\sum_{k_y} e^{-i\,m\,k_y\, a_y}~ \psi_{n,k_y}$, one can represent the Hamiltonian $H$ [see Eq. (\ref{Eq3})] in  momentum space as $H=\sum_{k_y} H_{k_y}$, where

\begin{align}
&H_{k_y}=  -t_x\sum_{n=1}^{N-1} (\psi_{n+1,k_y}^\dagger \psi_{n,k_y}+ \text {H.c.}) \label{Eq4} \\
&-\sum_{n=1}^{N} \Big[ 2\,t_y \cos(k_y a_y+ 2\pi n a_x/\lambda)+\mu-2 t_x \Big]
\psi_{n,k_y}^\dagger \psi_{n,k_y} . \nonumber
\end{align}
This Hamiltonian $H_{k_y}$ exactly matches the Hamiltonian for the one-dimensional CDW modulated NW [see Eq. (\ref{Eq1})] upon the substitutions  $k_y \, a_y \rightarrow \alpha$ and $t_y \rightarrow V$. We note that now the entire 2D system decomposes into a set of $M$ independent 1D systems. The phase offset $\alpha$ plays the role of the momentum $k_y$. The amplitude of the CDW $V$ is replaced by the tunneling amplitude $t_y$. The period of the CDW modulation of the chemical potential is set by the strength of the applied magnetic field, given by $\lambda=h/eB\,a_y$. With these substitutions we can interpret the spectrum shown in Fig. \ref{fig02} as the dispersion ($E$ as function of $k_y$) of a two-dimensional electron gas in the QHE regime with pertinent gaps [\onlinecite{Oded1},\onlinecite{JK6}].  For the isotropic case $t_x=t_y$, we recover the standard Landau levels for the integer QHE.
 Finally, the $\nu$ bound states of the CDW modulated NW case are mapped to $\nu$ dispersive chiral QHE edge states, see also Fig. \ref{fig02}

\begin{figure*}[htb]
\begin{center} \begin{tabular}{ccc}
\epsfig{figure=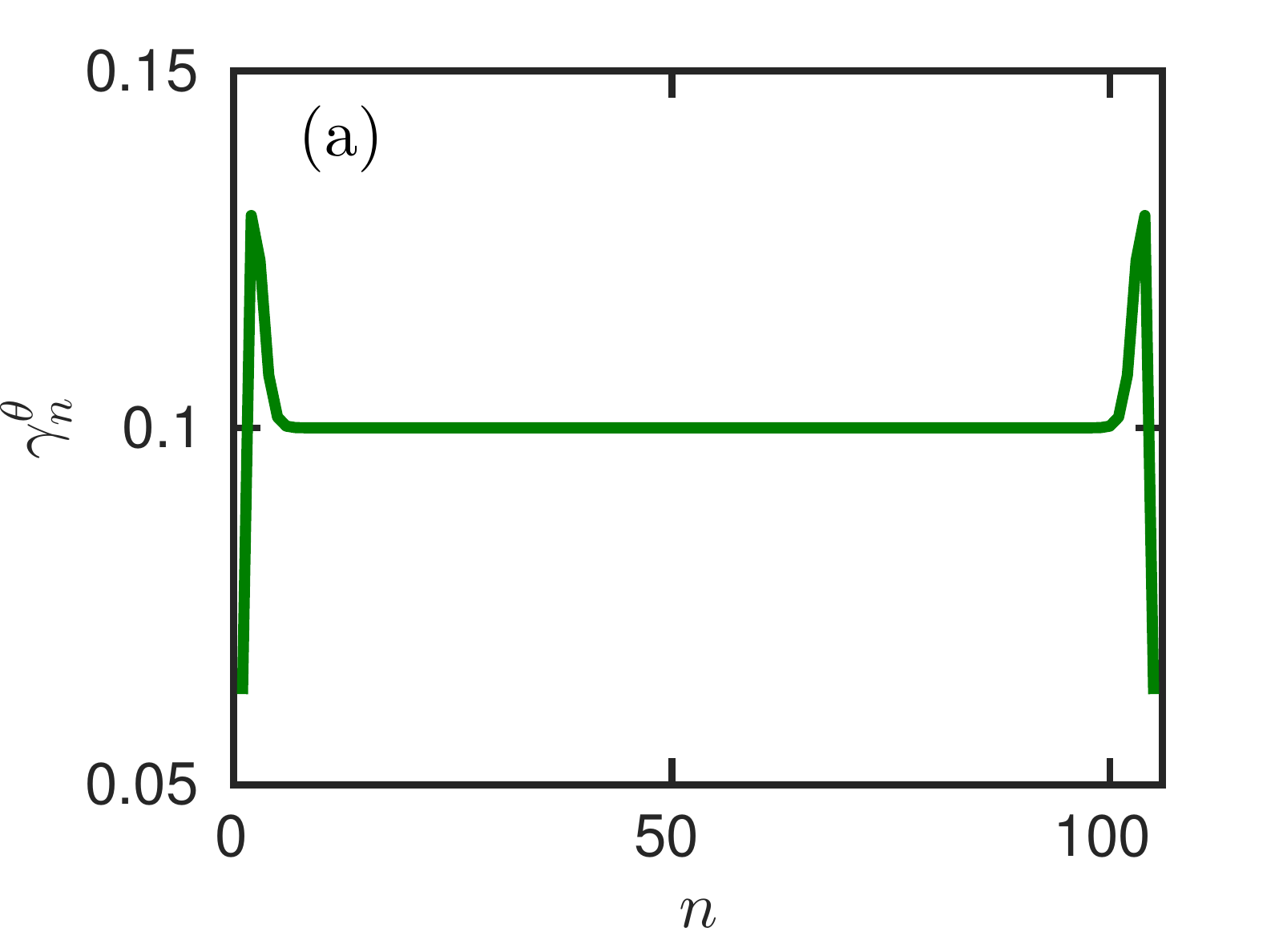,width=2.2in,height=1.75in,clip=true} &\hspace*{-0.cm}
\epsfig{figure=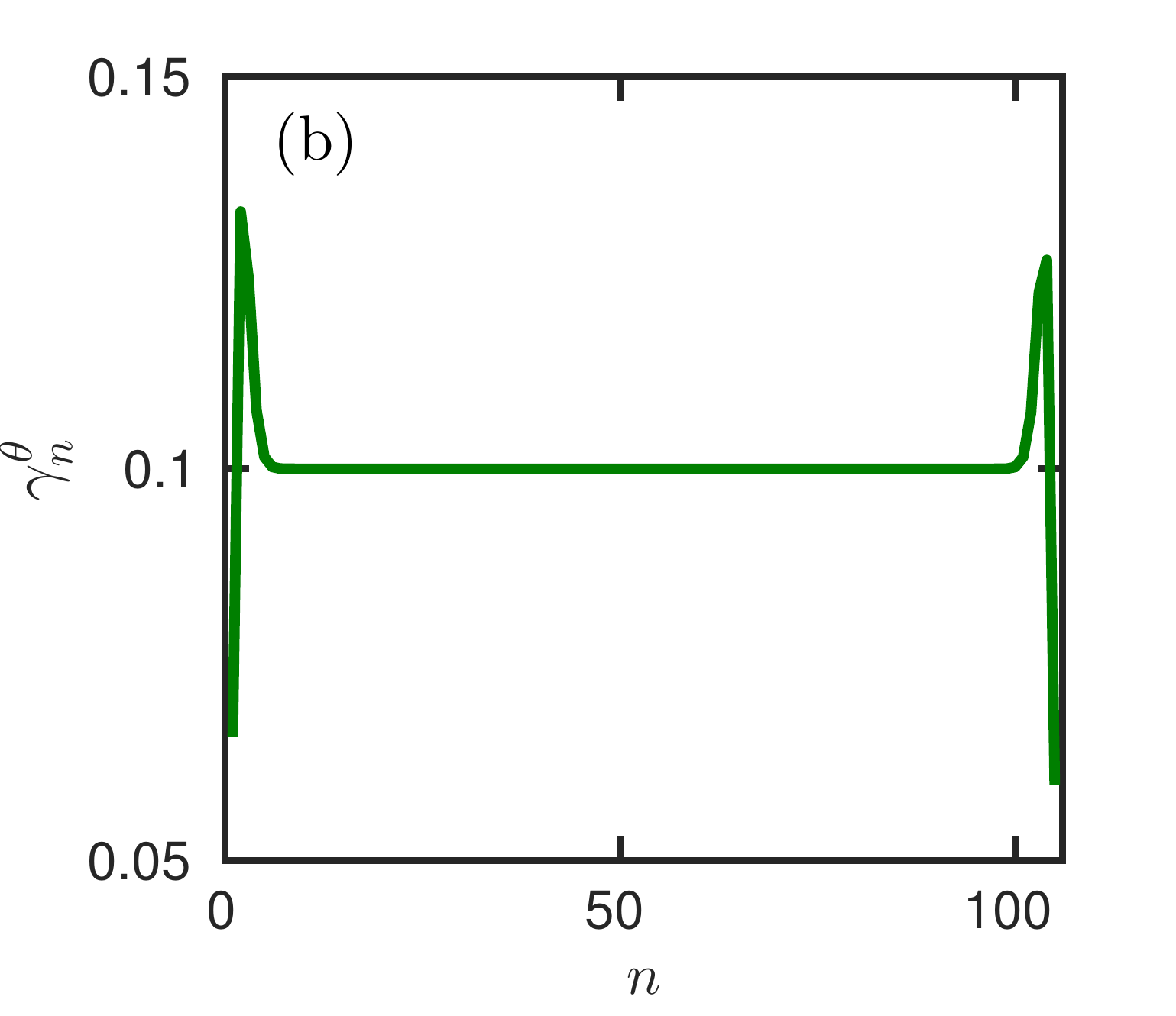,width=2.2in,height=1.8in,clip=true} &\hspace*{-0.cm}
\epsfig{figure=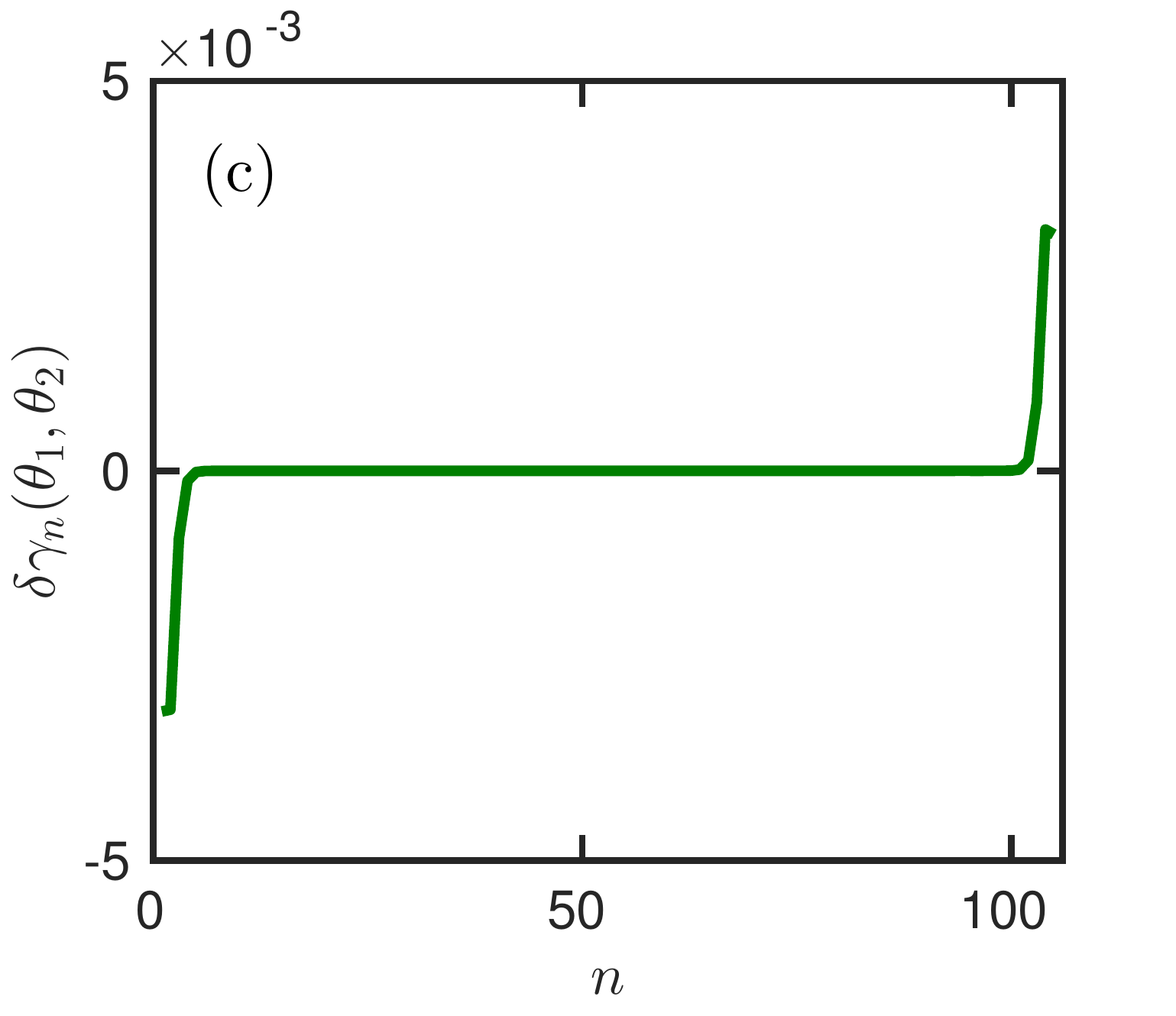,width=2.2in,height=1.8in,clip=true}\\
\epsfig{figure=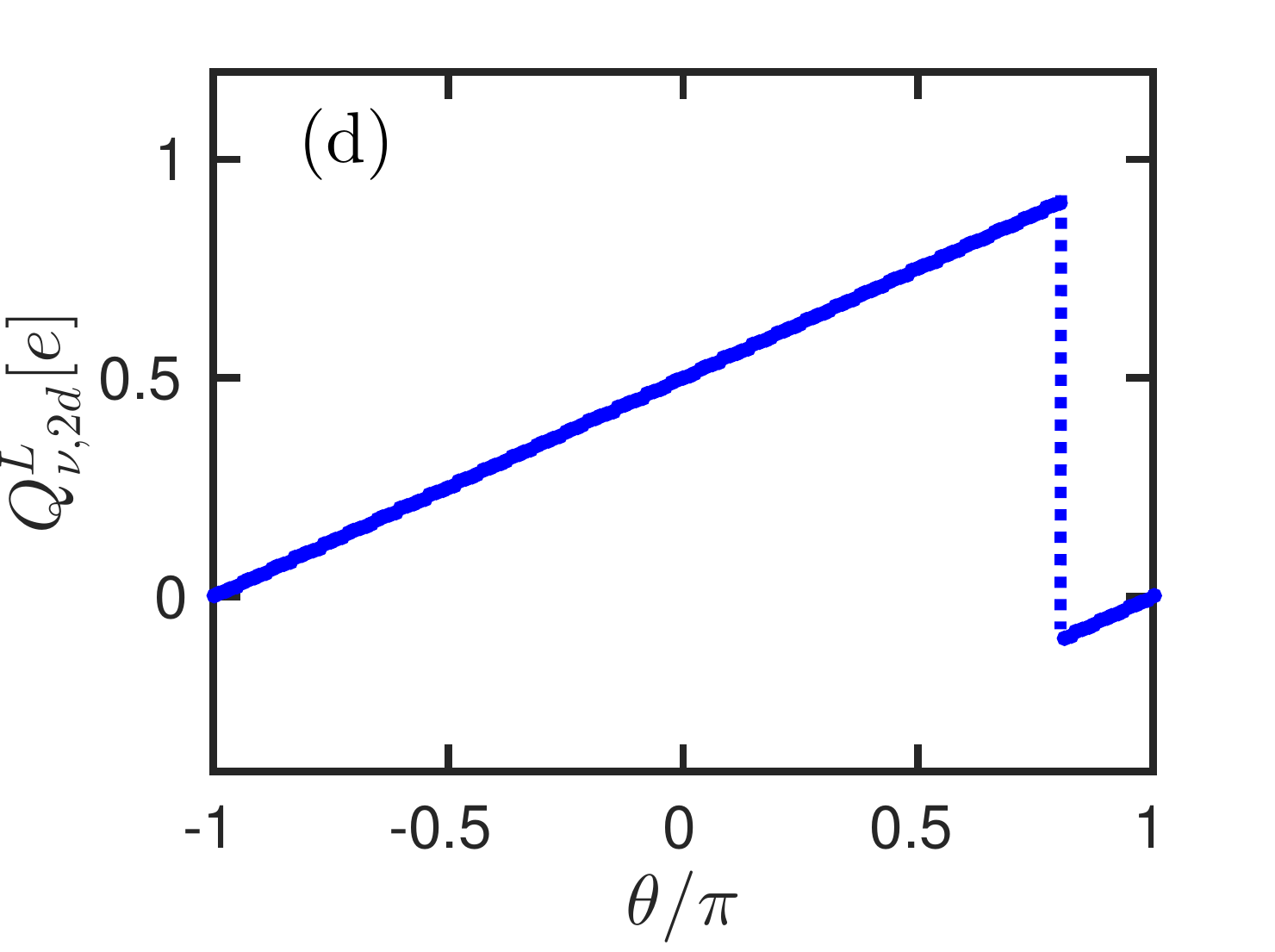,width=2.3in,height=1.8in,clip=true} &\hspace*{-0.cm}
\epsfig{figure=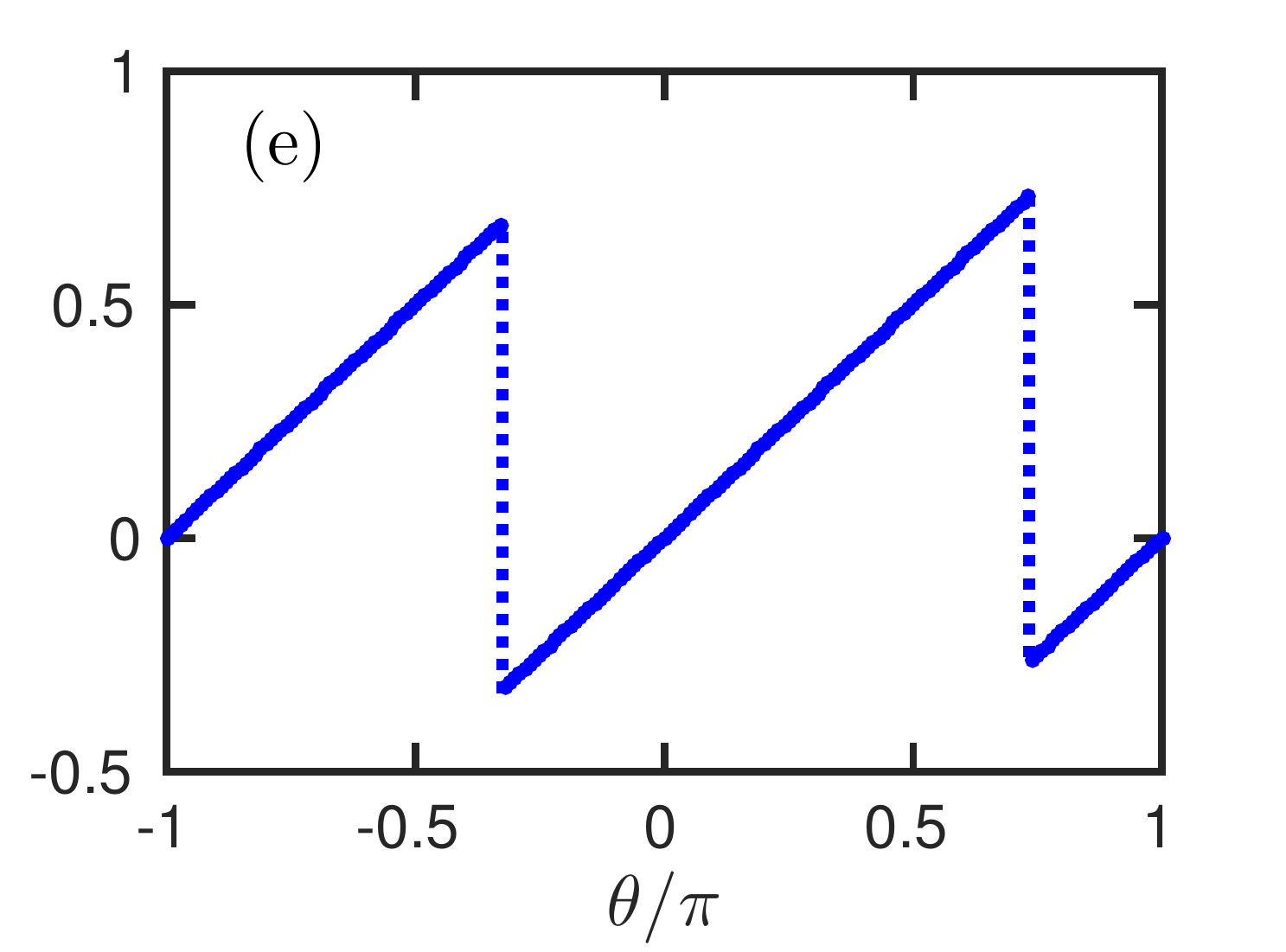,width=2.2in,height=1.8in,clip=true} &\hspace*{-0.cm}
\epsfig{figure=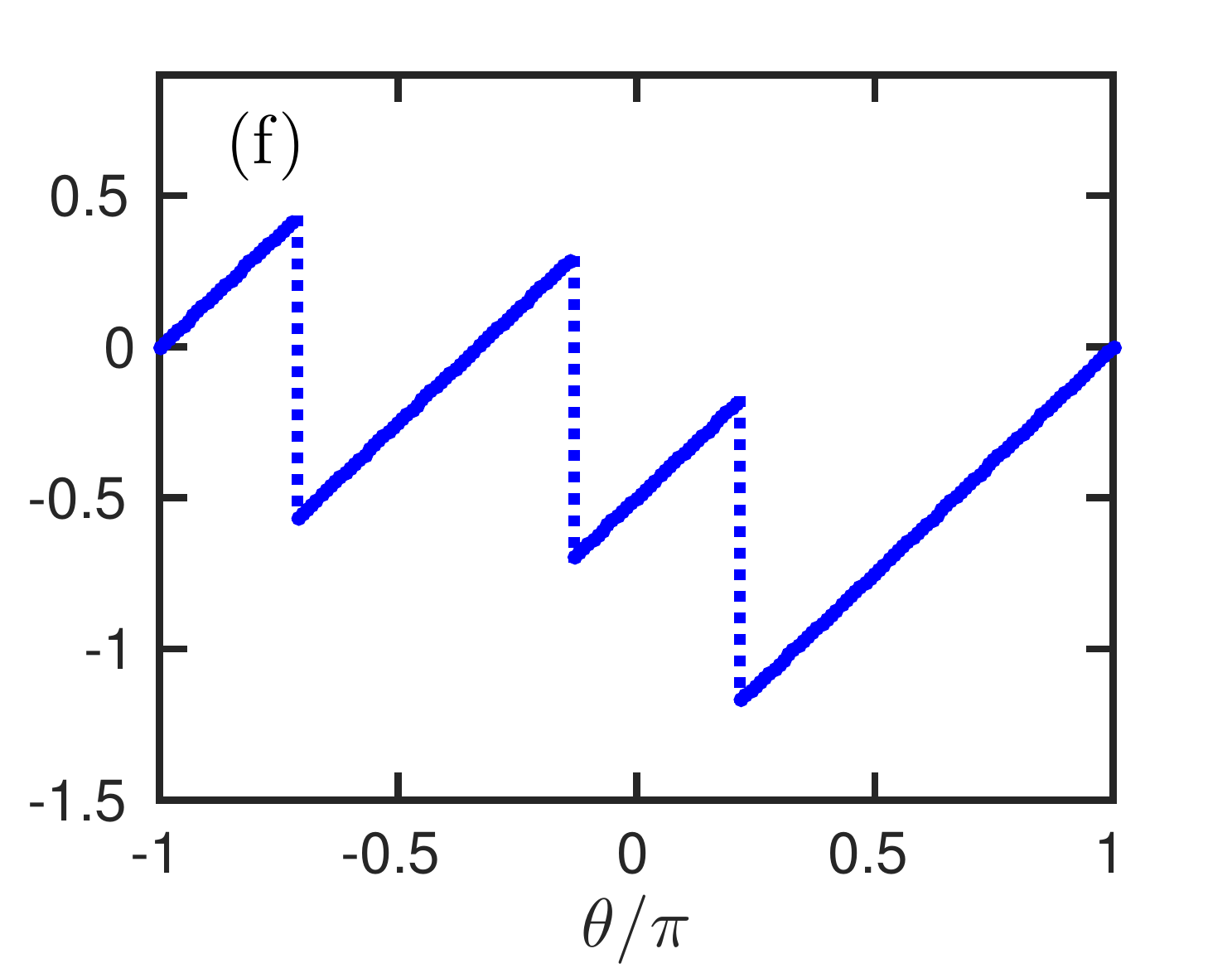,width=2.2in,height=1.8in,clip=true} 
\end{tabular} \end{center}
\caption{ Local particle density $\gamma_n ^\theta$ for the QHE setup with chemical potential being tuned inside the first bulk gap ($\mu/t_x=-0.1$) as a function of the position along the NW for two values of the flux phase: (a) $\theta_1=0$ and (b) $\theta_2=0.9$. In  panel (c), we plot the difference in the local particle densities $\delta \gamma_n(\theta_1,\theta_2)=\gamma_n^{\theta_1}- \gamma_n ^{\theta_2}$ calculated at these two values of $\theta$. The bulk value of $\gamma_n^\theta$ is universal and depends only on the number of filled bands. As a result, $\gamma_n^\theta$  is sensitive to the flux only at the boundaries, which results in the dependence of the FBCs on $\theta$. The FBCs $Q_{\nu,2d}^{R,L}$ change linearly as a function of the flux. The linear slope $c_\nu=\nu e/2\pi$ is perfectly quantized and the FBCs show all the four salient features described in Sec. \ref{sec2}, now as a function of flux phase, $\theta$.  The chemical potential $\mu$ is tuned inside the (d) first [$\mu/t_x =-0.1$, $\nu=1$], (e) second [$\mu/t_x =0.5$, $\nu=2$], (f) third [$\mu/t_x =1$, $\nu=3$] bulk gap.
The parameters are fixed as $t_y/t_x=0.6$, $\lambda/a_x=10$, $N=105$, $M=20$,  $n_1=20$, and $n_2=1$.
 }
\label{fig07}
\end{figure*}

\section{FBC in QHE system with flux}
\label{sec4}

In foregoing section, we established the connection between the 1D CDW-modulated NW and an array of tunnel-coupled NWs in the QHE regime. Now, we are in the position to introduce the FBC for the 2D system. In the 1D case, the FBC depends on the phase offset $\alpha$, which maps to the momentum $k_y$ in the 2D setup. While $\alpha$ can be controlled experimentally and  tuned to  different but fixed values, in any finite 2D system all bulk states with different momentum $k_y$ compose the ground state, and, thus, they all contribute to the FBCs. Hence, we should revisit the concept of  FBCs in 2D. 

First, we  introduce the particle density,  $\gamma_{n,m}$ at the site $n$ of the $m$th NW defined as 
\begin{align}
 \gamma_{n,m}= \langle \psi_{n,m}^\dagger \psi_{n,m} \rangle.
 \label{Eq9}
\end{align}
Here the expectation value is calculated in the ground state of the system.  In  the 1D case, even far away from the NW ends, the charge density is non-uniform over the CDW period $\lambda$. In the 2D case, however, the charge density is uniform in the bulk due to the global translational invariance, see Fig. \ref{fig07}.  Thus, in 1D  we were forced to compensate for this non-uniformity in the definition of the FBCs by introducing the second cut-off $n_2$ in the profile function $f^{R,L}_n$, see Eq. (\ref{frl}). In contrast to that, in the 2D setup we can work with $n_2=1$ in the profile function $f^{R,L}_n$ and just make sure that $n_1 a_{x}$ exceeds the localization length of the QHE edge states. These considerations allow us to introduce the FBCs  $Q_{\nu,2d}^s$ for the 2D setup as follows:
 \begin{align}
 Q_{\nu,2d}^s= \sum_{(n,m)=(1,1)}^{(N,M)} f^s_{n} \,(e \, \langle \psi_{n,m}^\dagger \psi_{n,m} \rangle- \rho_\nu).
 \label{Eq11}
\end{align} 
Here, $s=R,L$ labels the FBC at the right and left boundary of the system, respectively. The index $\nu$ indicates the position of the chemical potential inside the $\nu$th bulk gap with the bulk charge per site  given by $ \rho_\nu$. For simplicity,  we work with the same profile function $f^s_{n}$ for all NWs, see Eq. (\ref{frl}). We have checked that this choice does not affect our results.

Next, we impose periodic boundary conditions along the $y$ direction, giving rise to a cylinder topology, see Fig. \ref{fig06}. In addition, we add an external flux  $\Phi$. The flux is created by an additional external magnetic field $B'$ aligned along the $x$ axis. The corresponding vector potential ${\bf A'}$ is chosen to be along the $y$ axis, ${\bf A'}=(B' R /2)\,\hat y$, where $R = M a_y/2\pi$ is the radius of the cylinder. The total flux penetrating the cylinder is defined as $\Phi=\int {\bf A'}\cdot {\rm d} {\bf l}= \pi R^2 B'$.

The corresponding Hamiltonian in the tight-binding model is defined as
\begin{align}
&H^{\theta}=\Big[-t_x\sum_{(n,m)=(1,1)}^{(N-1,M)} \psi_{n+1,m}^\dagger \psi_{n,m}\nn&-t_y \, e^{i\,\theta/M}\sum_{(n,m)=(1,1)}^{(N,M)} e^{i \,2\,\pi\,n\,a_x/\lambda}\psi_{n,m+1}^\dagger  \psi_{n,m} \Big]+\text{H.c.}\nn
&-(\mu-2\,t_x) \sum_{(n,m)=(1,1)}^{(N,M)}\psi_{n,m}^\dagger \psi_{n,m},
\label{Eq8}
\end{align}
 where $\theta/M =(e/\hbar)\int {\bf A'}\cdot {\rm d}{\bf l}=e\,B'\,R\,a_y/(2\hbar)=2\, \pi\,\Phi/(\varphi_0\,M)$  is the Peierls phase that the electrons acquire by tunneling between two neighboring NWs. Here,  $\varphi_0=h/e$ is the flux quantum. For simplicity of notations, we identify the $(M+1)$th NW with the first NW. 

By applying the Fourier transformation and introducing the momentum $k_y$, we again find that the 2D Hamiltonian can be represented as a sum of independent 1D Hamiltonians in momentum space, $H^{\theta}=\sum_{k_y} H^{\theta}_{k_y}$, where
\begin{align}
H^{\theta}_{k_y} = H_{k_y} \big(k_y a_y \to k_y a_y + \theta/M \big).
\label{Eq81}
\end{align}
By changing the flux $\Phi$ through the cylinder, one can effectively shift the momentum $k_{y}$.

If the system is periodic along the $y$ direction (as assumed),  the particle density $\gamma_{n,m}^{\theta}$ is independent of the NW index $m$,  $\gamma_{n,m}^{\theta}\equiv \gamma_{n}^{\theta}$, see Fig. \ref{fig07}.  Here, we have introduced the dependence of the particle density on the  flux phase $\theta$. Of course, this dependence is only significant at the boundaries of the system, as the bulk value of the particle density, $ \rho_\nu/e$, is a constant determined by the position of the chemical potential inside the $\nu$th bulk gap, see Fig. \ref{fig07}. 

Numerically, one can easily show that the FBCs  $Q_{\nu,2d}^s (\theta)$ depend linearly on the flux phase $\theta$, see Fig. \ref{fig07}.  The slopes at the right and left boundary are opposite,
\begin{align}
 Q_{\nu,2d}^R=- c_\nu\theta+C^R; \ \ Q_{\nu,2d}^L= c_\nu \,\theta+C^L,
 \label{QRL}
 \end{align}
and depend solely on the fact that the chemical potential is positioned inside the $\nu$th bulk gap, $c_\nu=e\,\nu/2\pi$. To insure the $2\pi$ periodicity of the FBC, again there must be $\nu$ jumps of size $\pm e$ as the flux phase $\theta$ changes by $2\pi$. This feature is again ensured by the non-universal piecewise constant functions $C^{R,L}$. 

The linear dependence of the FBCs on $\theta$ can also be understood analytically by using the mapping to the CDW-modulated NW. By applying the Fourier transformation to the definition of the FBCs $Q^s_\nu (\theta)$, we arrive at
$Q^s_{\nu,2d} (\theta)=\sum_{k_y}Q^s_{k_y} (\theta)$, where $Q^s_{k_y}$ is the FBC defined for an effectively 1D Hamiltonian $H^{\theta}_{k_y}$. Making use of Eq. (\ref{QR}) for the FBCs in 1D systems, in which we replace $\alpha$ by $k_y a_y + \theta/M$, we arrive at Eq. (\ref{QRL}). We note that, as the sum runs over all $M$ values of quantized momentum $k_y$, {\it i.e.},  over the entire Brillouin zone, such that $\sum_{k_y} k_y a_y=0$, while $\sum_{k_y} \theta/M=\theta$. This confirms the linear dependence of the FBCs on the flux phase $\theta$ with the universal slope $c_\nu=e\,\nu/2\pi$.

\begin{figure*}[tb]
\begin{center} \begin{tabular}{ccc}
\epsfig{figure=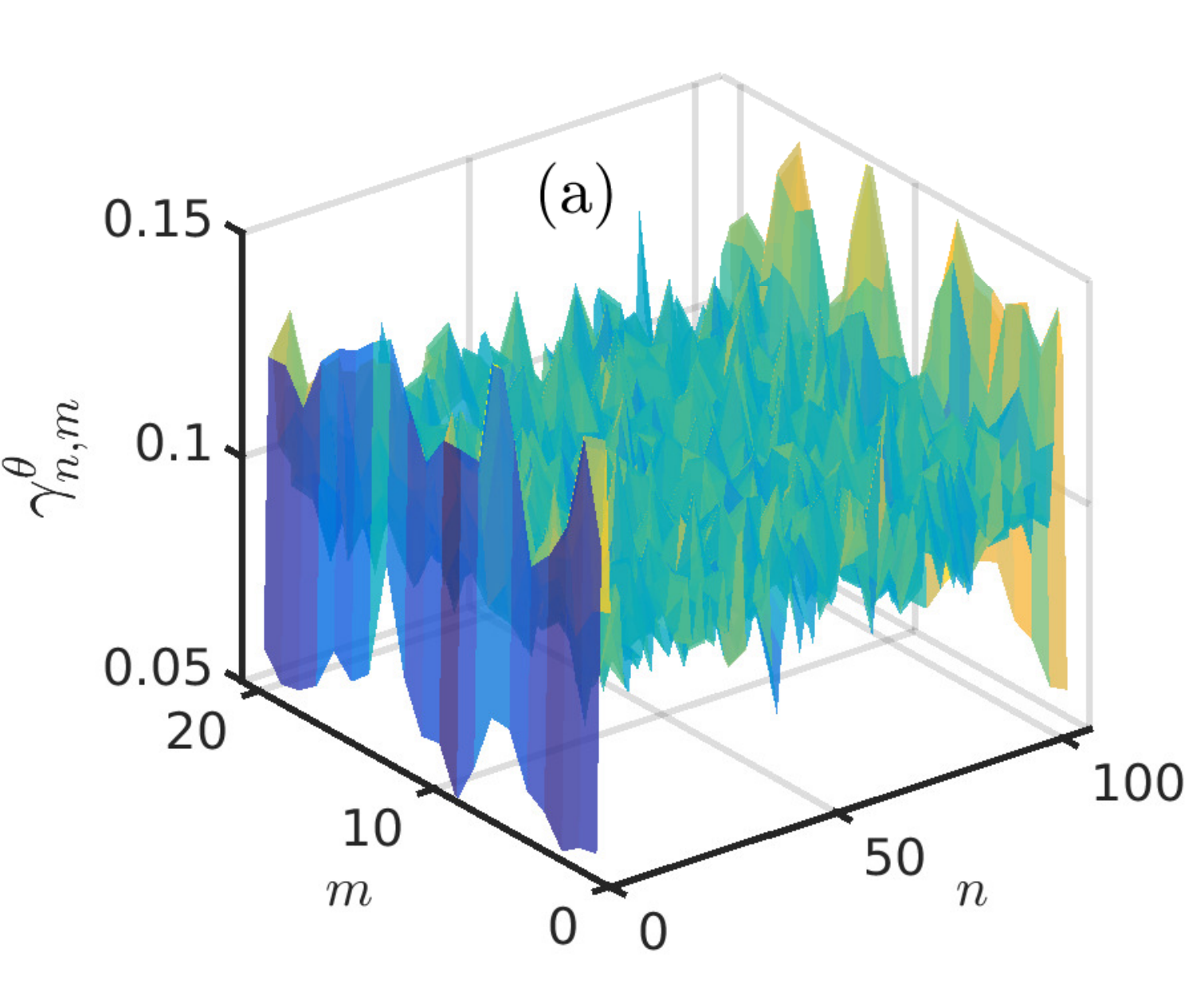,width=2.2in,height=1.8in,clip=true} &\hspace*{-0.cm}
\epsfig{figure=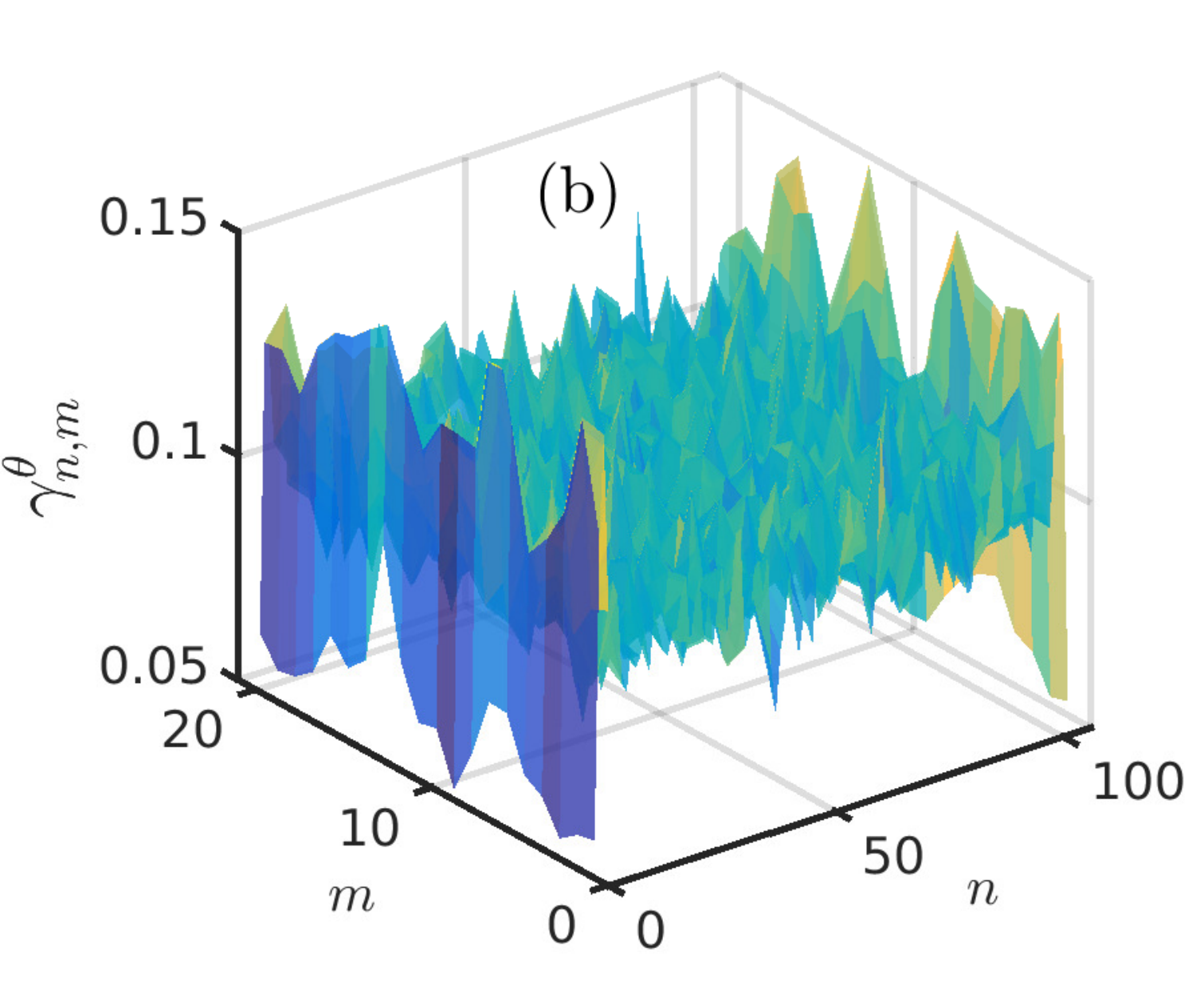,width=2.2in,height=1.8in,clip=true} &\hspace*{-0.cm}
\epsfig{figure=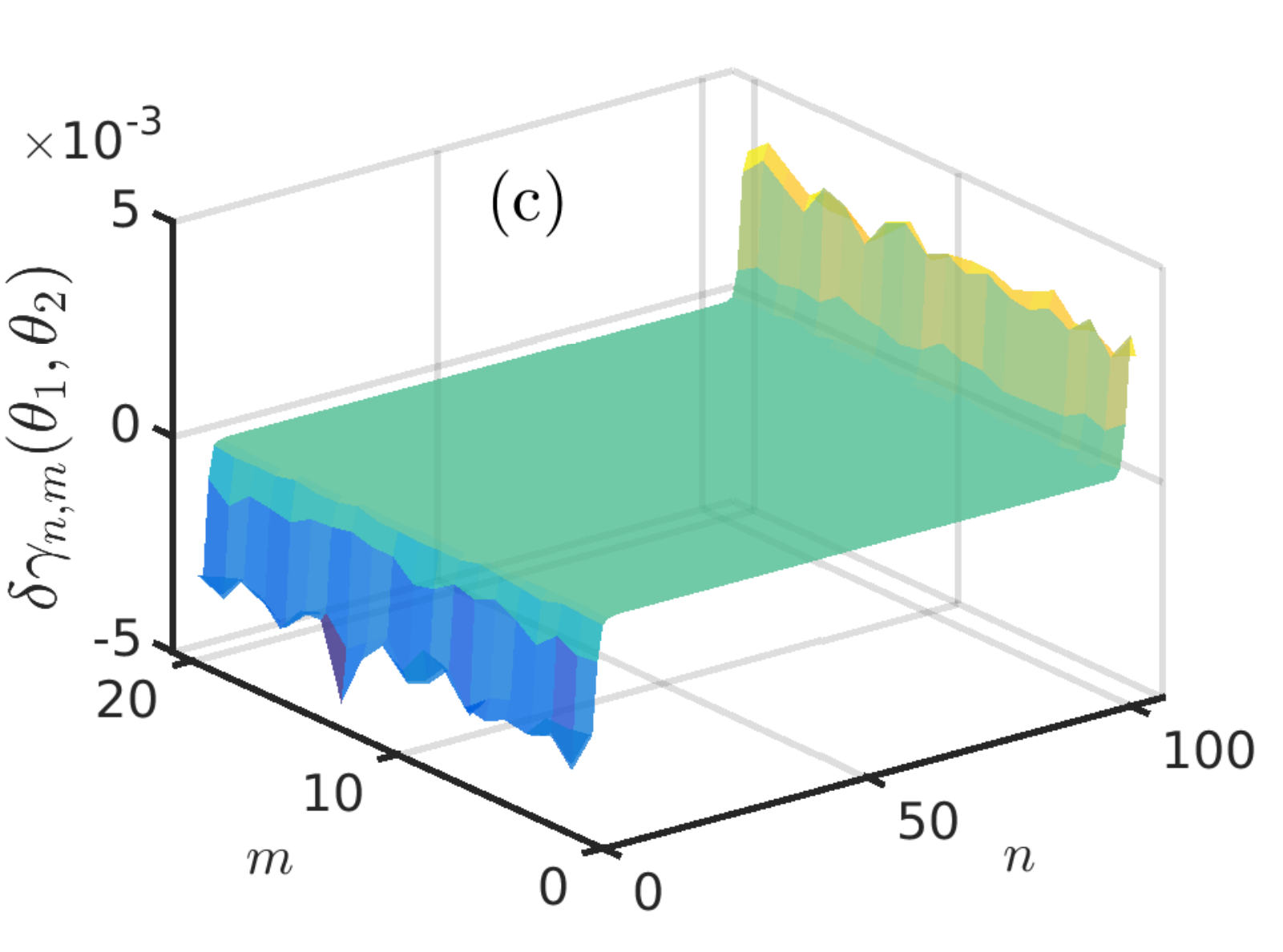,width=2.2in,height=1.8in,clip=true} \\
\epsfig{figure=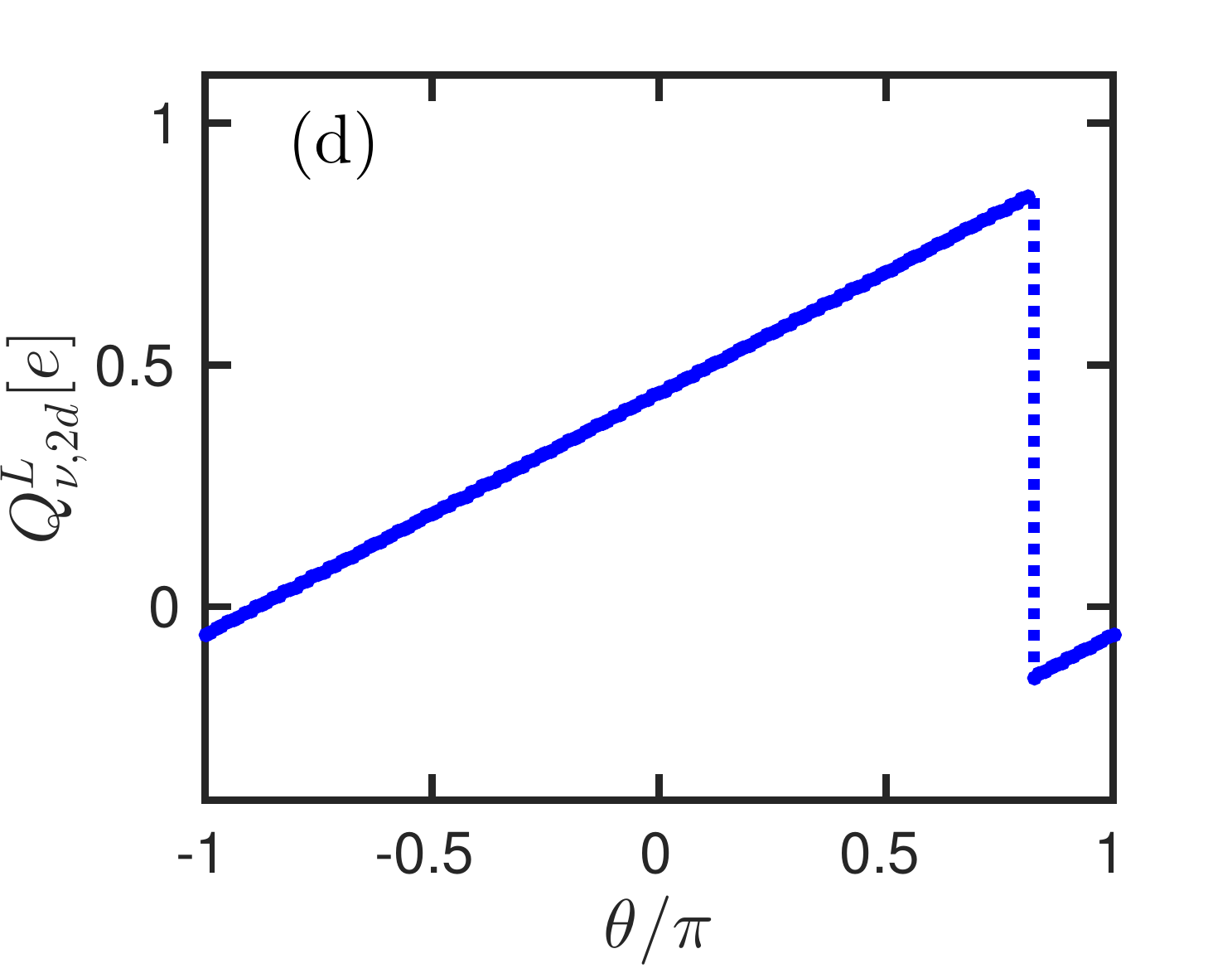,width=2.2in,height=1.8in,clip=true} &\hspace*{-0.cm}
\epsfig{figure=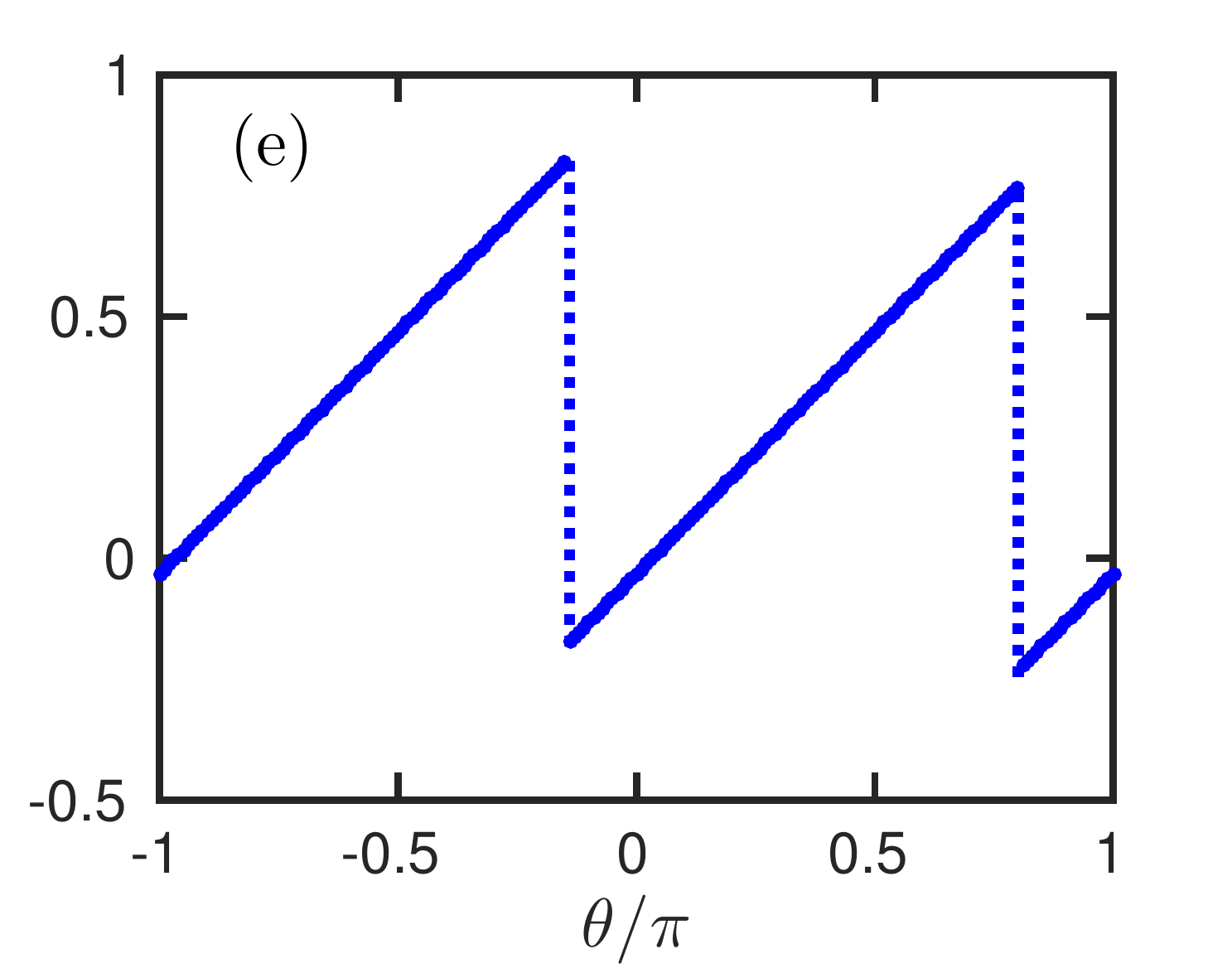,width=2.1in,height=1.8in,clip=true} &\hspace*{-0.cm}
\epsfig{figure=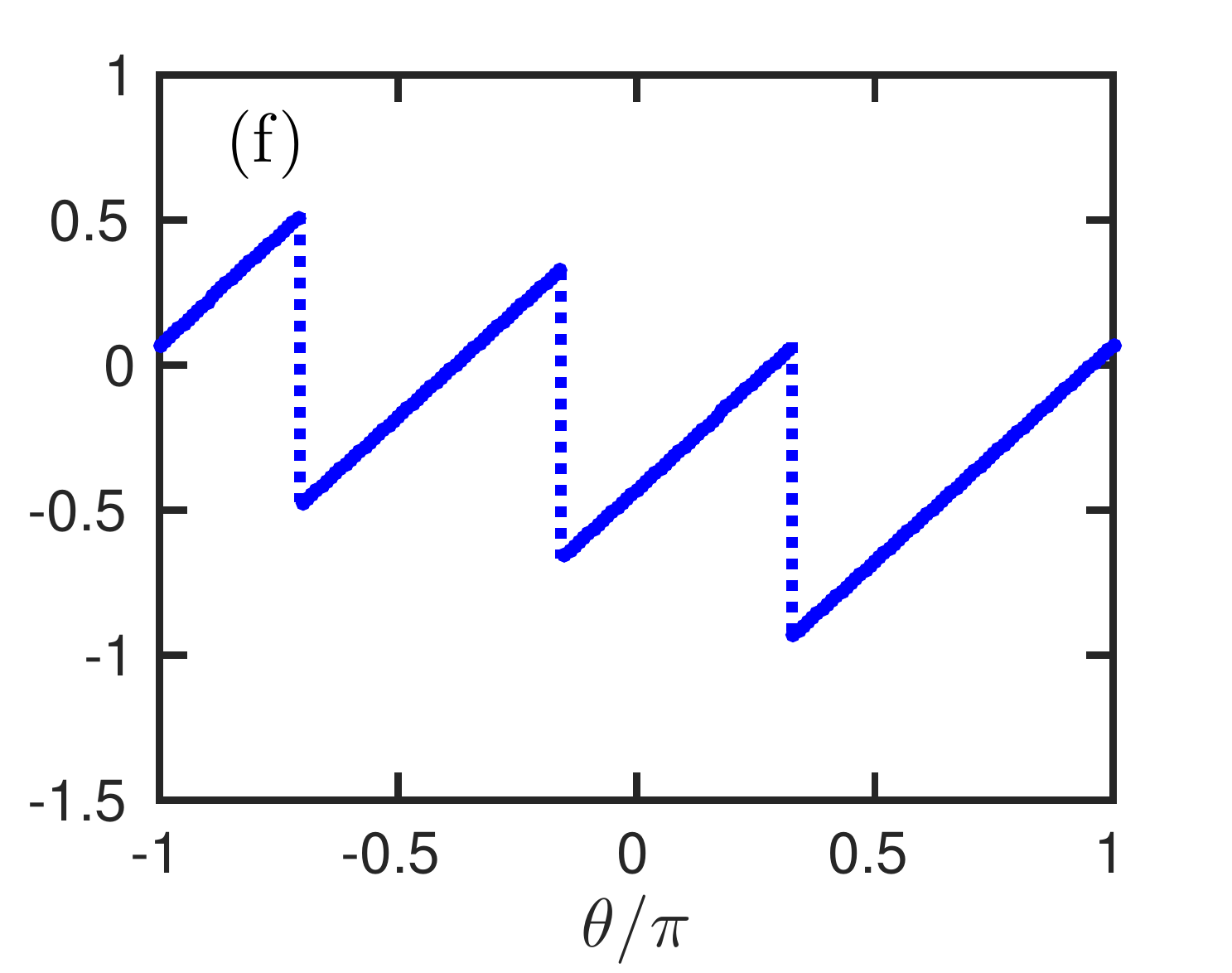,width=2.1in,height=1.8in,clip=true}
\end{tabular} \end{center}
\caption{The same as in Fig. \ref{fig07}, however, in the presence of strong disorder modeled by on-site fluctuations of the chemical potential with standard deviation $\sigma_d/t_x=0.1$. The spatial distribution of the particle density $\gamma_{n,m}^{\theta}$ [(a) $\theta_1=0$ and (b) $\theta_2=0.9$] is non-uniform even in the bulk due to disorder. However,  as in the clean case, the particle density depends on $\theta$ only at the boundaries of the system. Indeed, as seen from panel (c), the difference 
$\delta \gamma_{n,m}(\theta_1,\theta_2)=\gamma_{n,m}^{\theta_1}- \gamma_{n,m} ^{\theta_2}$ is zero in the bulk but finite at the boundaries.
(d-f) The left FBC ($Q_{\nu,2d}^L$) shows exactly the same dependence on the magnetic phase $\theta$ as in the clean case [compare with Fig. \ref{fig07} (d-f)]. This demonstrates the robustness of the linear slope of FBCs to disorder.
}
\label{fig08}
\end{figure*}

\section{FBC and   disorder in QHE regime}
\label{sec5}

In previous sections, for our analytical arguments, the periodicity of the system was crucial in order to establish the linear dependence of the FBCs $Q^s_{\nu,2d}$ on the flux phase $\theta$. However, in realistic samples, disorder can be substantial and break this periodicity so that $k_y$ is no longer a good quantum number. Thus, it is crucial to check the stability and universality of the linear slopes in the FBCs as a function of flux also in the presence of disorder, where the disorder is allowed to be very general, in particular to be present in the entire 2D system including the edges. This can be easily done numerically in the tight binding model, where we add an onsite disorder term $\mu_{n,m}^{dis}$ characterized by a normal distribution, where the mean value, without loss of generality, is fixed to zero, while the standard deviation $\sigma_d$ controls the distribution of the values of $\mu_{n,m}^{dis}$,
\begin{align}
H_{dis}=\sum_{(n,m)=(1,1)}^{(N,M)} \mu_{n,m}^{dis} \,\psi_{n,m}^\dagger \psi_{n,m}.
\label{Eq15}
\end{align}
The  full tight-binding Hamiltonian takes the form, $H_{tot}=H^\theta+H_{dis}$, where $H^\theta$ is given by Eq. (\ref{Eq8}). We again first calculate numerically the particle densities $ \gamma_{n,m}^\theta$ for different values of magnetic flux, see Fig. \ref{fig08}.  In contrast to the clean case, $ \gamma_{n,m}^\theta$ are not constant anymore even in the bulk of the system, {\it i.e.}, away from the sample boundaries. However, the mean value of $\gamma_{n,m}^\theta$ stays close to the clean bulk limit value $ \rho_\nu/e$. Again, the largest deviations from $\rho_\nu/e$ are observed at the boundaries of the system. Surprisingly, if one focuses on the changes in the particle densities $\gamma_{n,m}^\theta$ as the flux phase $\theta$ is adjusted for the same configuration of disorder, one notices that $\gamma_{n,m}^\theta$ in the bulk of the system is independent of the magnetic flux value. By calculating the difference of particle densities  for two different values of flux phases $\theta_1$ and $\theta_2$,  $\delta \gamma_{n,m}{(\theta_1,\theta_2)} = \gamma_{n,m}^{\theta_1} -\gamma_{n,m}^{\theta_2}$, we find that $\delta \gamma_{n,m}{(\theta_1,\theta_2)}$ takes non-zero values only at the boundaries, see Fig. \ref{fig08}(c). In comparison with the clean case, the particle density $\gamma_{n,m}^{\theta}$ is non-uniform along the boundary as the translation-invariance is broken by disorder. However, this local redistribution of the particle density along the boundary does not effect the FBCs. Importantly, also in the presence of strong disorder, $Q_{\nu, 2d}^{L,R}$ reproduces a linear dependence on the flux phase $\theta$ [see Fig. \ref{fig08} (d-f)] and Eq. (\ref{sigma}) is valid. We note that, in what follows, we are interested in the part of the FBC, $Q_{\nu,2d}^{R,L}$, that depends on the flux phase $\theta$. Thus, even if in the presence of strong disorder the average particle density in the bulk can deviate from the clean case value $\rho_\nu$ used in the definition of the FBC [see Eq. \ref{Eq11}], it does not play any role in further discussions, in which we will be interested only in differences in the FBCs, $\delta Q_{\nu,2d}^{R,L}(\theta_1,\theta_2)=Q_{\nu,2d}^{R,L}(\theta_1)-Q_{\nu,2d}^{R,L}(\theta_2)$. Obviously, the constant $\rho_\nu$ used in Eq. \ref{Eq11} does not play any role as it cancels exactly in the expression for $\delta Q_{\nu,2d}^{R,L}(\theta_1,\theta_2)$. However, this allows us to explain a slight offset between the values of the FBCs obtained in the clean [see Fig. \ref{fig07}] and disordered [see Fig. \ref{fig08}] cases. 

In addition, similarly to the bound states in the 1D case described above, we note that it is the chiral edge states in the QHE regime that are responsible for the finite jump in the FBCs $Q_{\nu,2d}^{R,L}$.  This jump is always in integer steps of the elementary charge $e$ and can be understood as follows. The summation in the definition of $Q_{\nu,2d}^{R,L}$ [see Eq. \ref{Eq11}] runs over a length that is larger than the typical localization length (in $x$ direction) of the edge states. Therefore, each  filled edge state below the chemical potential  contributes fully to the boundary charge, {\it i.e.}, 
$e \sum_{(n,m)=(1,1)}^{(N,M)} f^s_{n}  \, \langle \psi_{n,m}^\dagger \psi_{n,m} \rangle_{\rm{filled}} =e$.  Thus, if a filled edge state crosses the chemical potential as a function of $\theta$, there is an integer jump in units of $e$ in $Q_{\nu,2d}^{R,L}$. Away from such crossing points, $Q_{\nu,2d}^{R,L}$ is a smooth linear function of $\theta$, see Fig. \ref{fig08}. Conversely, this also means that the edge states do not contribute to the linear {\it slope} of the FBCs and the slope comes solely from  boundary contributions of extended bulk states which change as function of flux. We emphasize that while strong disorder can result in  states that are fully localized in the system, numerically, we  observe a substantial amount of  bulk states that are extended over the whole system including  both boundaries. 
Finally, fully localized states in the spectrum are independent of the flux and do not contribute to the slope of the FBCs either.

Generally, the FBCs for the 2D system in the QHE regime exhibit all the four salient features which we have discussed in earlier sections. These features are also independent of the details of the profile functions $f_n^{L,R}$. All these findings highlight the robustness of the obtained results. The value of the linear slope $c_\nu$ is {\it universal} (independent of system parameters) and perfectly quantized in units of $e/2\pi$, $c_\nu = \nu e/2\pi$. All these suggests that  this slope can be used as a topological invariant for the system. Importantly, in contrast to many other topological invariants such as winding numbers or Chern numbers which rely on the periodicity of the system and, thus, can be calculated only in the clean case, the topological invariant  $c_\nu$ is well-defined even in the presence of strong disorder in the whole system. In the next section we will establish the connections between $c_\nu$ and the quantized Hall conductance. 

\begin{figure*}[t]
\begin{center} \begin{tabular}{cc}
\epsfig{figure=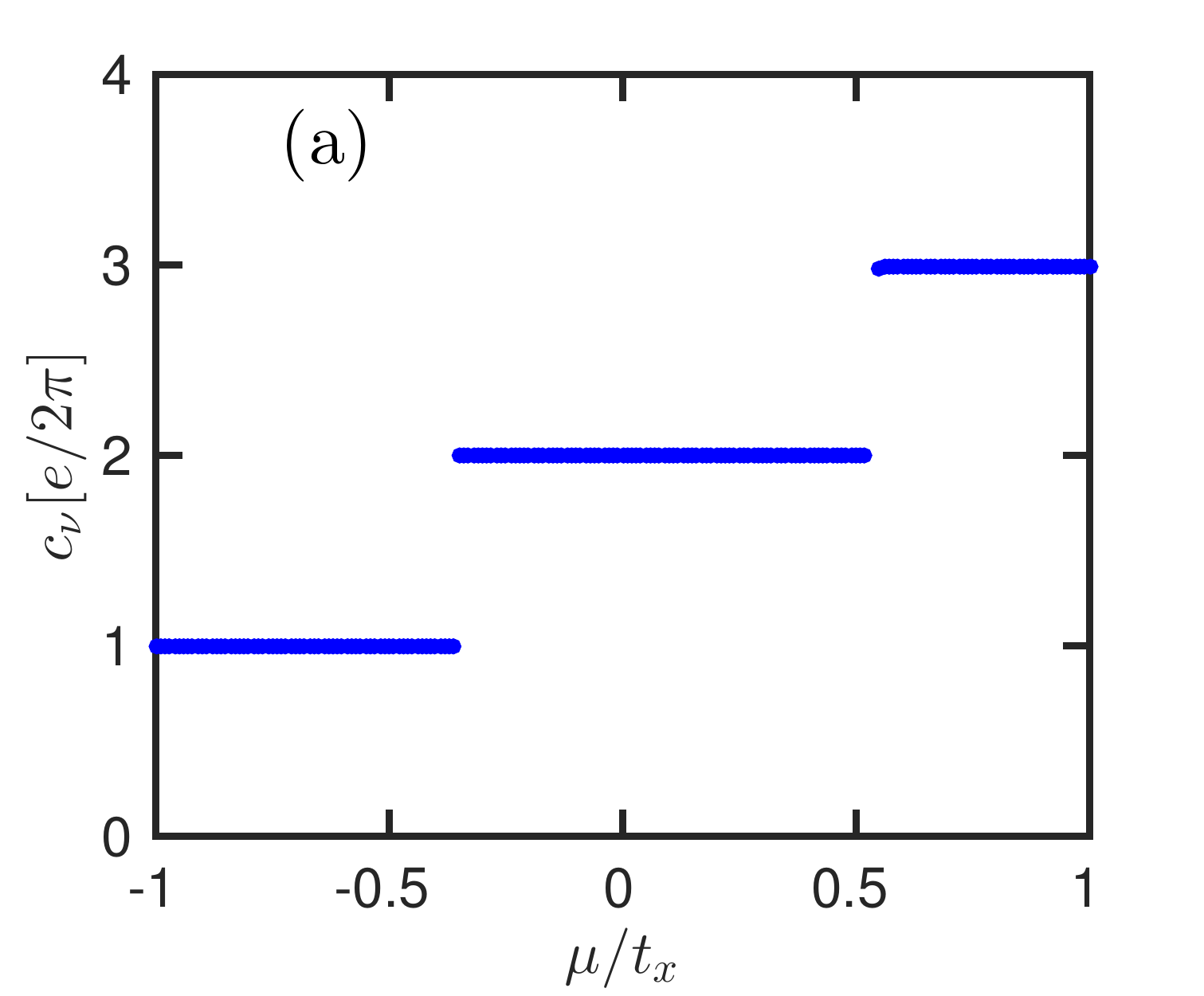,width=2.4in,height=1.9in,clip=true} &\hspace*{0.5cm}
\epsfig{figure=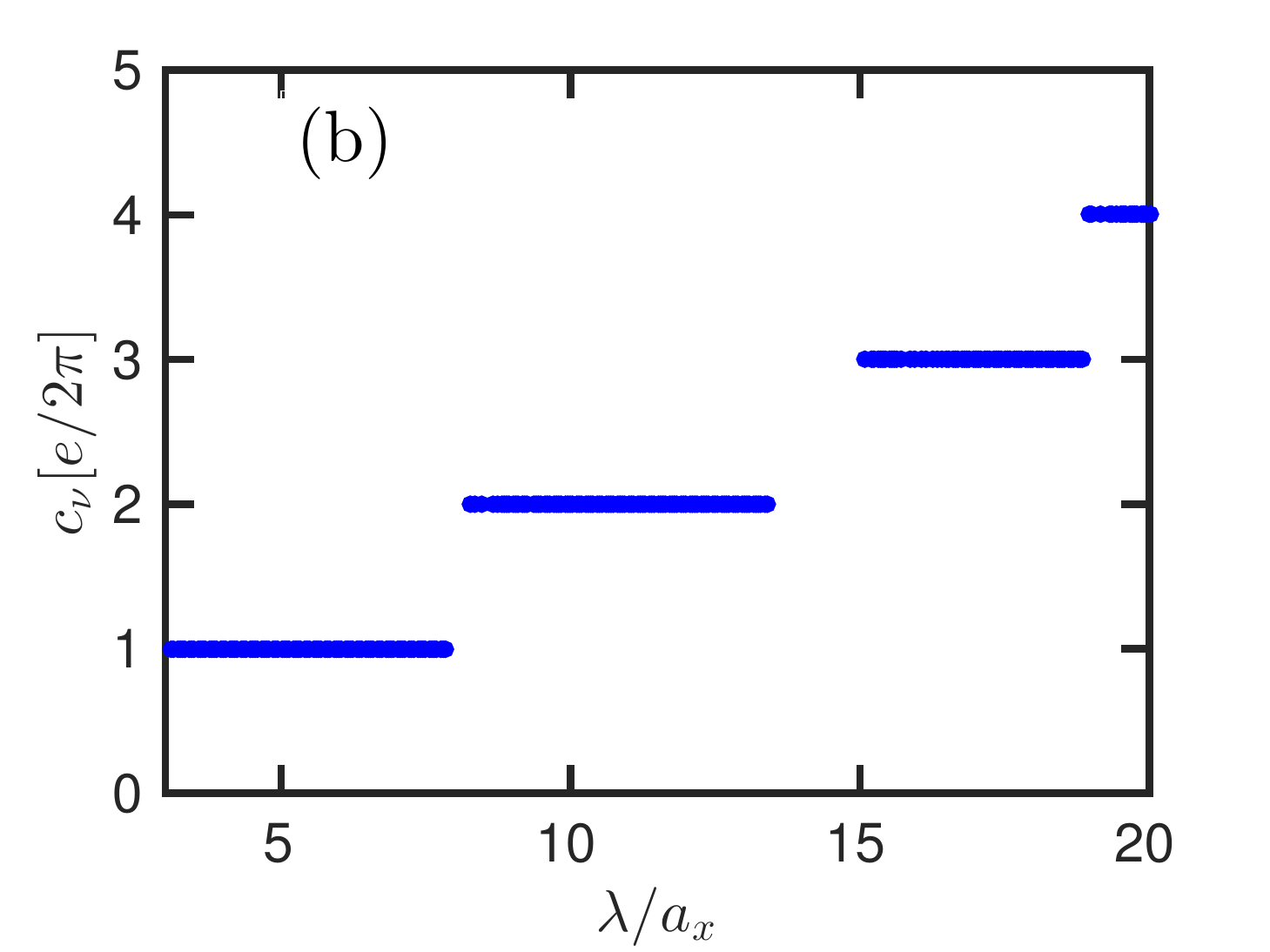,width=2.4in,height=1.9in,clip=true}
\end{tabular} \end{center}
\caption{The slope $c_\nu$ [in units of $e/2\pi$] (a) as function of the position of the chemical potential $\mu$ and (b) as a function of $\lambda=h/e B a_y$. We focus on the isotropic regime with $t_y/t_x=1$ and fix either (a)  the magnetic field $\lambda/a_x=10$ or (b) the position of the chemical potential $\mu/t_x=0$. The remaining parameter values  are chosen  as in Fig. 7. The quantization of the slope $c_\nu$ results in quantized plateaus in the Hall conductance $\sigma_{xy}$.  }
\label{fig11}
\end{figure*}

\section{FBC and Hall conductance}
\label{sec6}

In this section, we show that the FBC allows one to address explicitly the Hall conductance of the QHE system. For this we need to connect the FBC to the Hall current. We start by introducing the total charge of a small patch of  area $\cal A$ located at the system boundary [see  Fig. \ref{fig06}] as $Q_{{\cal A},2d}= Q_{\nu,2d}^L+Q_{{\cal A},2d}^b$, where $Q_{{\cal A},2d}^b$ is the bulk contribution defined as $Q_{{\cal A},2d}^b = \rho_\nu {\cal A}/a_x a_y$. We note that $Q_{{\cal A},2d}^b$ is independent of $\theta$. 

The continuity equation, $ \partial \rho_{2d}/ \partial t+{\bf \nabla}\cdot {\bf j}=0$, connects the charge density, $\rho_{2d}(x,y)=e  \langle \psi (x,y)^\dagger \psi (x,y) \rangle$ [continuum version of $\gamma_{n,m}$ given in Eq. (\ref{Eq9})], with the current density ${\bf j}(x,y)$ in standard notation. Next, we integrate the equation over the patch ${\cal A}$ and use the Gauss theorem $\int_{\cal A} {\bf \nabla}\cdot {\bf j}\,  d{\cal A}=\int_{\partial {\cal A}} {\bf j}\cdot d{\bf S}$, connecting the volume integral over the area ${\cal A}$ to the surface integral over the closed patch  $\partial {\cal A}$. Here, the surface differential $d{\bf S} $ is a vector pointing normal to the boundary of the patch  $\partial {\cal A}$. Thus, the continuity equation can be rewritten as
\begin{align}
 \frac{d Q_{{\cal A},2d}}{d t}+\int_{\partial {\cal A}} {\bf j}\cdot  d{\bf S}=0,
 \label{Eq12}
\end{align}  
where  $Q_{{\cal A},2d}=  \int_{\cal A} \, \rho_{2d}\, d{\cal A} $.
For the patch located at the boundary of the system, only current along the $x$ axis crossing the boundary between the patch and the bulk of the system, $I_x$, contributes to the integral. This allows us to define the total current $I_x =\int_{\partial {\cal A}} {\bf j}\cdot d {\bf S}$. From the continuity equation Eq.(\ref{Eq12}), we get $I_x =-\Dot Q_{{\cal A},2d}$.

Next, we change the FBCs in time by changing the flux $\Phi$ through the cylinder. The bulk contribution $Q_{{\cal A},2d}^b$ stays constant and the change in the total charge is only due to the change in the FBC, $Q_{\nu,2d}^L$. Using Eq. (\ref{QRL}), we obtain $\Dot Q_{{\cal A},2d} =\Dot Q_{\nu,2d}^L= c_\nu \, \Dot \theta= 2 \pi \, c_\nu \,\dot \Phi\,/\varphi_0$. According to the Faraday law, the change of flux in time generates the electromotive force $\mathcal{E}_y$ acting along the $y$ axis, $\dot \Phi=-\mathcal{E}_y$. Combining the two expressions for the change of the FBC,  we arrive at the following relation between the current $I_x$ and the electromotive force $\mathcal{E}_y$,
\begin{align}
I_x= -\frac {d Q_{\nu,2d}^L}{dt}=  2 \pi \, c_\nu \,\frac{\mathcal{E}_y}{\varphi_0}.
\label{Eq13}
\end{align}

As a result, the Hall conductance $\sigma_{xy}$ is given by
\begin{align}
\sigma_{xy}=\frac{I_x}{\mathcal{E}_y}= \frac{2 \pi\, c_\nu}{\varphi_0}.
\label{sigma}
\end{align}
Using the values of the linear slope  $c_\nu= \frac{e\,\nu}{2\pi}$ found above both analytically and numerically, we find that the conductance  takes the form $\sigma_{xy}= \frac{\nu\, e^2}{h}$, which are the quantized values of the integer QHE.

Remarkably, the linear slope of the FBCs takes quantized value that leads to the  quantized Hall conductance.  We also compute numerically the dependence of $c_\nu$ (and thus of the Hall conductance $\sigma_{xy}$) on the position of the chemical potential as well as on $\lambda$ (which is inversely proportional to magnetic field $B$), see Fig. \ref{fig11}. The slope $c_\nu$ is an integer multiple of $e/2\pi$ inside a given gap. As one increases $\mu$ or $\lambda$, more bands get filled and $c_\nu$ changes by $e/2\pi$ as one of the bulk bands crosses the chemical potential, see Fig. \ref{fig02}. The plateaus in $c_\nu$ correspond to plateaus in the  Hall conductance $\sigma_{xy}$ and they are stable against disorder as shown in previous section. Therefore, our approach gives an alternative way to microscopically understand  the Hall conductance and its robust quantization. As seen in previous sections, the slope of the FBCs have contributions from all occupied bands. Thus, also the Hall conductance gets contributions from all occupied bands, except from the edge states which, however, are responsible for the discontinuous jump from one plateau to the next. 
 
 We note that our approach is also valid for finite temperatures  as long as the temperature stays smaller than the energy distance from the chemical potential to the nearest bulk band, see Appendix \ref{ap3}. As soon as the temperature is high enough to thermally excite electrons from localized edge states to extended bulk states (or vice versa), the FBCs cannot be defined properly anymore. As a result, the linear dependence of the FBCs on the flux breaks down. Hence, as the temperature increases, the Hall plateaus begin to shrink before disappearing eventually.

We also would like to emphasize the advantages of the approach presented here over the Laughlin argument [\onlinecite{Laughlin}]. First, the change in the flux $\Phi$ does not need to be an integer multiple of the flux quantum $\varphi_0$ as assumed in Laughlin's argument [\onlinecite{Laughlin}], but instead can take any value. Second, and more important, our derivation is valid also in systems with strong disorder, whereby the disorder can be present in the whole sample including the boundaries. The linear dependence of the FBCs on the magnetic flux holds also in this case, which highlights the remarkable stability of the quantized values against disorder. This stability suggests that the slope $c_\nu$ of the FBCs plays the role of a topological invariant which is well defined even in the presence of strong disorder. Finally, we note that our derivation is  valid for any position of the chemical potential inside the bulk gap, see Fig. \ref{fig11}.

\section{Conclusions and Outlook}
\label{sec7}

We have studied FBCs occurring in one-dimensional nanowires with periodically modulated chemical potential as well as in two-dimensional electron gases in the presence of a perpendicular magnetic field in the integer QHE regime. In the clean limit, these two systems can be mapped onto each other. In both systems, the FBCs are linear functions of the phase offset (1D case) or of the magnetic flux in the cylinder topology of the Laughlin setup (2D case). This linear slope $c_\nu$ depends only on the number of filled bulk bands but not on the precise position of the chemical potential inside these bands. The slope is universal and quantized in units of $e/2\pi$ and, moreover, is also extremely robust against disorder. Interestingly, $c_\nu$ is determined solely by bulk bands, while the bound states in 1D or the chiral edge states in 2D are responsible for the jumps in the FBCs, which are quantized in units of $e$. We have shown that all these features
are robust against disorder, and thus one  can consider $c_\nu$ as a topological invariant that is well-defined even in the presence of strong disorder.

In addition, we have shown that the direct consequence of quantized values of the slope $c_\nu$ is the quantization of the Hall conductance. Our derivation is performed for the Laughlin cylinder setup and, thus, can be tested experimentally in the Corbino disk geometry.  As only the bulk states are responsible for the finite slope  $c_\nu$, we conclude that the Hall current is  carried by extended bulk states. The FBCs and their change as function of phase offset in NWs or of flux in Corbino disks can be tested experimentally by making use of, for example, single electron transistors [\onlinecite{Amir1}, \onlinecite{Amir2}] as charge sensors. As an outlook, it would be interesting to generalize our approach to the Hall bar geometry.

\acknowledgements
This work was supported by the Swiss National Science Foundation (SNSF) and NCCR QSIT. This project received funding from the European Union’s Horizon 2020 research and innovation program (ERC Starting Grant, grant agreement  No 757725). 

\appendix
\section{Effective Hamiltonian for NW in higher order perturbation theory in $V$}
\label{ap1}
In this Appendix, we calculate explicitly the effective Hamiltonian describing the coupling between right and left movers at the Fermi surface in the case of higher-order resonances $k_F = \nu \pi/\lambda$ in the perturbative regime with $V\ll t_x$, as specified in the main part.
Generally, the only non-zero matrix elements of $H_{CDW}$  [Eq. (\ref{CDW})] in  momentum space, $M_{k_1,k_2}$, are the ones that connect two states with the momentum difference $2\pi/\lambda$,
\begin{align}
M_{k_1,k_2}&=\langle k_1|H_{CDW}|k_2 \rangle \nn&=- V\, e^{i\alpha}\delta_{k_1,k_2-2\pi/\lambda}- V\, e^{-i\alpha}\delta_{k_1,k_2+2\pi/\lambda}.
\label{a1}
\end{align}
\begin{figure}[b]
\includegraphics[width=0.85\columnwidth]{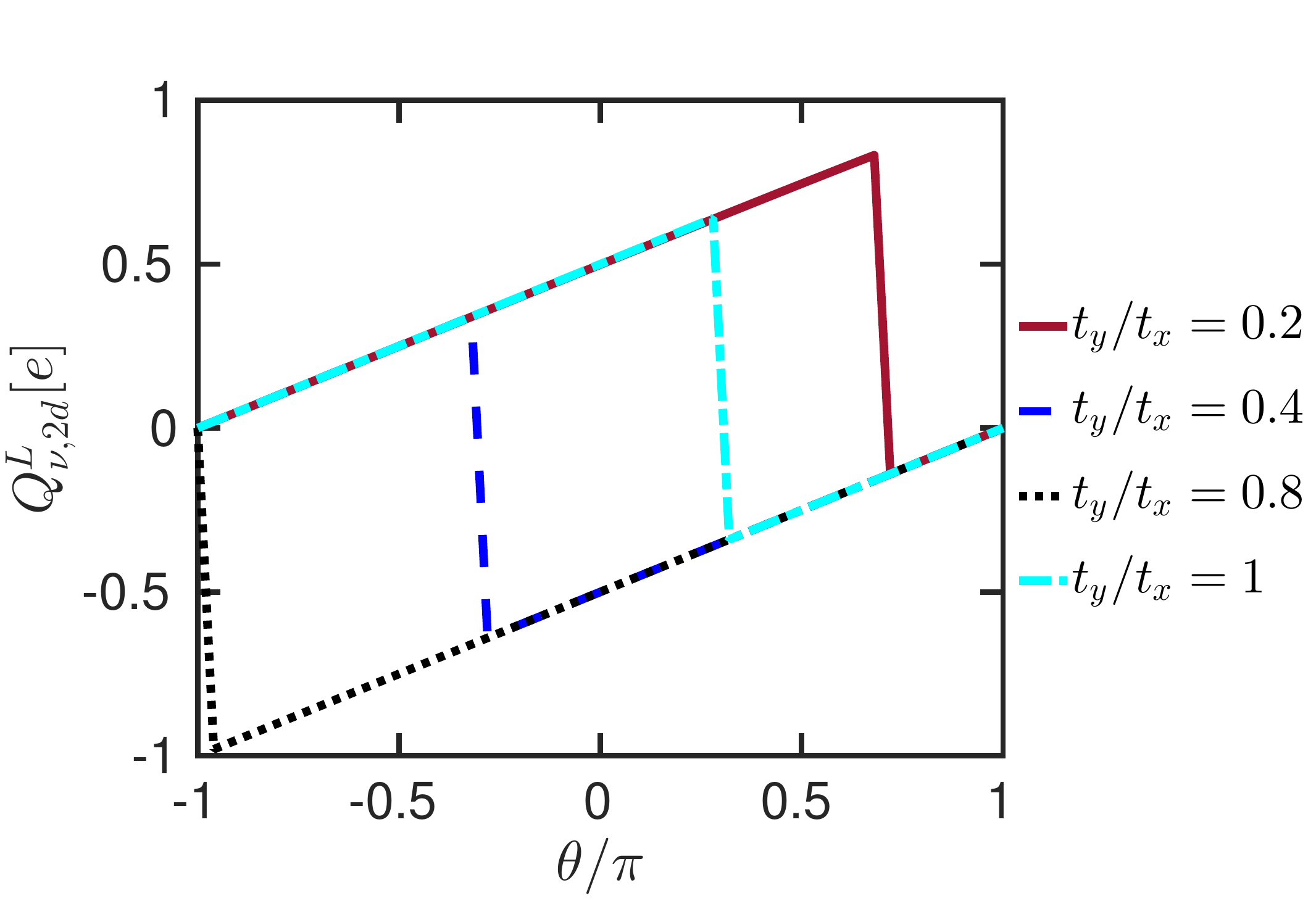}
\caption{The FBCs, $Q_{\nu, 2d}^{L}$, as a function of the flux phase $\theta$ for different values of the hopping amplitude between NWs: $t_y/t_x=1$ (cyan), 0.8 (black), 0.4 (blue), and 0.2 (red). The chemical potential $\mu$ is tuned into the first bulk gap ($\nu=1$): $\mu/t_x = -0.7, -0.5, 0,$ and 0.2, respectively. We observe the linear dependence of $Q_{\nu, 2d}^{L}$  on $\theta$. The
slope $c_\nu$ is quantized and independent of  whether the system is isotropic or not.  The remaining parameter values  are the same as in Fig. 7 of the main text.}
\label{fig10}
\end{figure}

\begin{figure*}[htb]
\begin{center} \begin{tabular}{cc}
\epsfig{figure=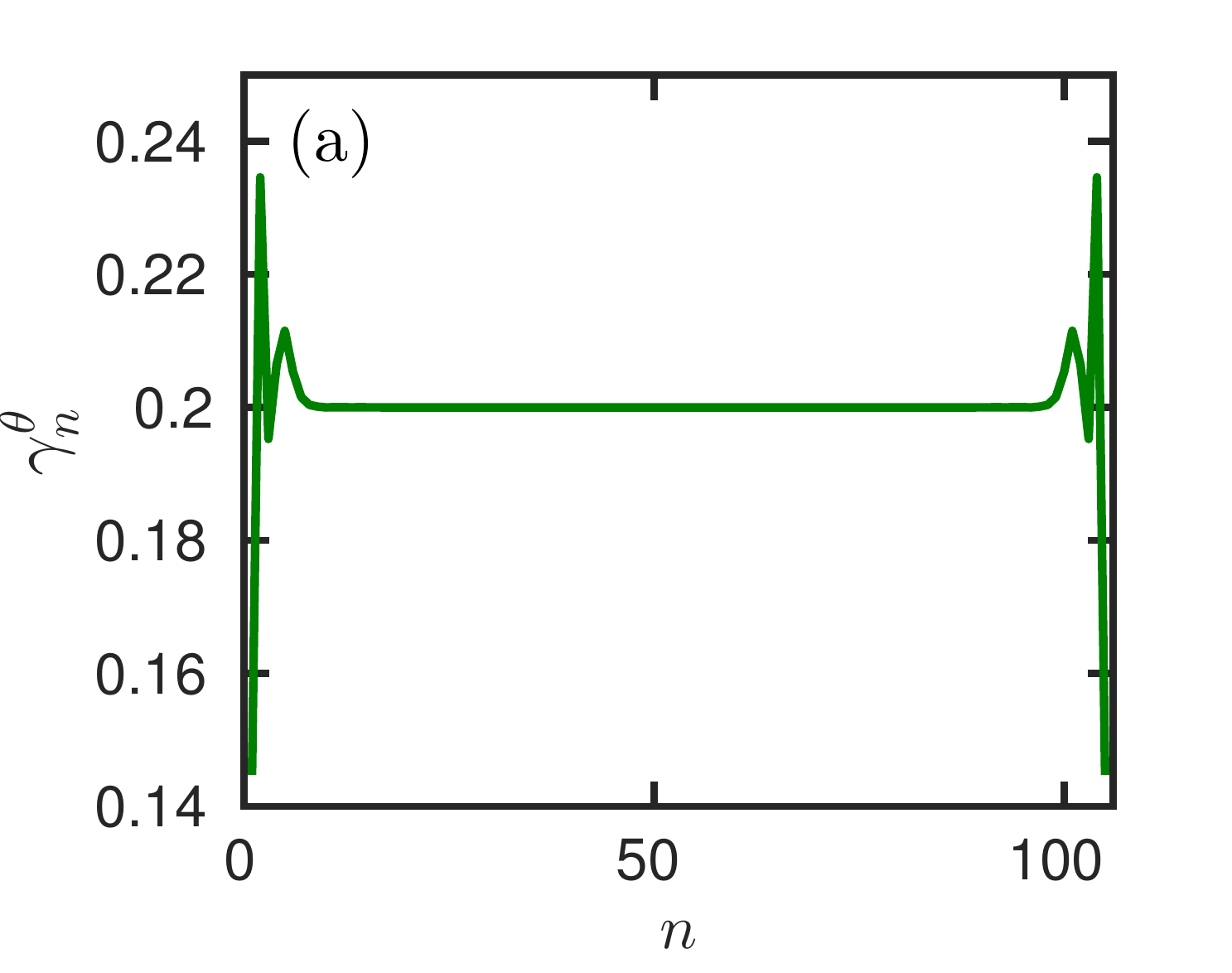,width=2.4in,height=1.9in,clip=true} &\hspace*{0.5cm}
\epsfig{figure=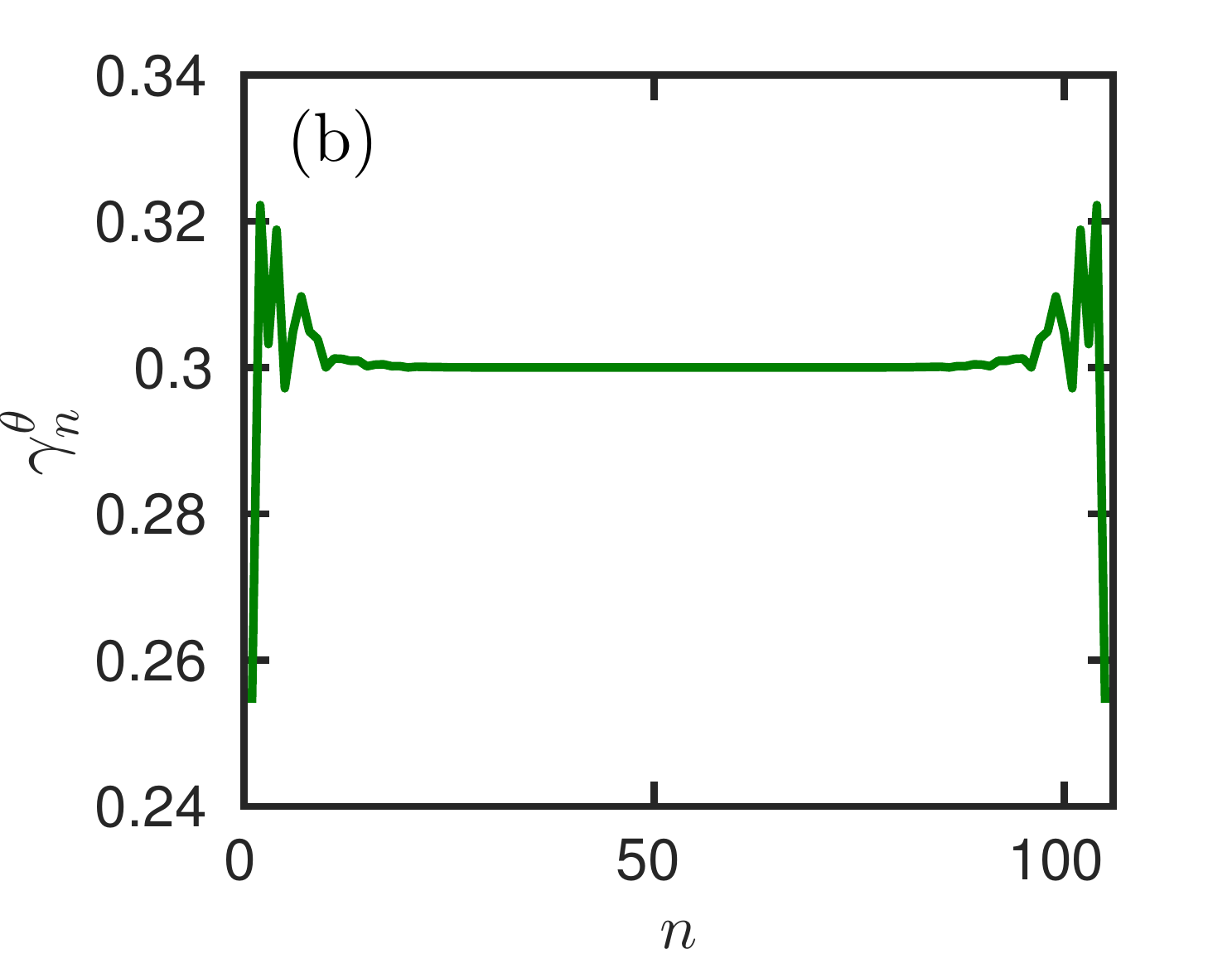,width=2.3in,height=1.9in,clip=true} 
\end{tabular} \end{center}
\caption{ Local particle density $\gamma_n^\theta$ as a function of position along the NW. The chemical potential lies (a) in the second ($\nu=2$) bulk gap with $\mu/t_x =0.5$ or (b) in the third ($\nu=3$) bulk gap with $\mu/t_x =1$. The particle densities are constant in the bulk but non-uniform at the boundaries. Other parameter values  are the same as in Fig. 7(a) of the main text. 
}
\label{fig09}
\end{figure*}

As a result, if $k_F = \nu \pi/\lambda$, the gap at the Fermi surface, $\Delta_g^{(\nu)}$, can be opened in the $\nu$th order perturbation theory [\onlinecite{M1},\onlinecite{M2}]. In this case, the effective Hamiltonian density in momentum space and in the basis $\tilde\psi=(R,L)$ is defined as
\begin{align} \label{qw1}
\mathcal{H}^{(\nu)}= \begin{pmatrix} 
\hbar\, v_F^{(\nu)}\,k &  {\bar\Delta}_g^{(\nu)}\\
{({\bar\Delta}_g^{(\nu)}})^* & -\hbar\, v_F^{(\nu)}\, k
\end{pmatrix},
\end{align}
with the Fermi velocity given by $v_F^{(\nu)}=2\,t_x\,a_x\, \sin(k_F\,a_x)/\hbar$. The matrix element ${\bar\Delta}_g^{(\nu)}$ connecting the right mover at the momentum $k_F$ and the left mover at the momentum $-k_F$ is found in the $\nu$th order perturbation expansion in $V$ as
\begin{widetext}
\begin{align}
{\bar\Delta}_g^{(\nu)} \equiv \Delta_g^{(\nu)}e^{i\,\nu\,\alpha}=\frac{M_{k_F,k_F-2\,\pi/\lambda}\cdots M_{-k_F-4\,\pi/\lambda,-k_F-2\,\pi/\lambda}M_{-k_F-2\,\pi/\lambda,-k_F}}{\prod_{q=1}^{\nu-1} [E^0_{-k_F}-E^0_{-k_F+2\,\pi\,q/\lambda}]}=\frac{V^{\nu} e^{i\,\nu(\alpha+\pi)}}{(4\,t_x)^{\nu-1}\prod_{q=1}^{\nu-1} \,\sin^2(k_F a_x q/\nu)},
\end{align}
\end{widetext}
where $E^0_k=2\,t_x[1- \cos(k\,a_x)]$ is the energy dispersion of the unperturbed Hamiltonian consisting only of the kinetic part. The spectrum of the effective Hamiltonian is given by $E_{\pm}=\pm [(\hbar\,v_F^{(\nu)}\, k)^2+(\Delta_g^{(\nu)})^2]$. The gap of the size $\Delta_g^{(\nu)}\sim V^{\nu}/ E_0^{\nu-1}$ is opened at the Fermi surface. Here, to simplify estimates, we introduced the characteristic energy $E_0$, which depends on the the position of the chemical potential and is of order of the Fermi energy.

 We note that Eq. (\ref{qw1}) obtained in the $\nu$th order of the perturbation theory maps back to the one considered in the main text ($\nu=1$) if one rescales $\alpha \rightarrow \nu \alpha$ and $V \rightarrow \Delta_g^{(\nu)}$. As a direct consequence, the number of bound states observed at any given energy inside the bulk gap as one tunes $\alpha$ from $-\pi$ to $\pi$ is also increased from one to $\nu$, see Fig. \ref{fig02}.

\section{FBCs in 2D models with different degree of anisotropy $t_y/t_x$} 
\label{ap3}
In this  Appendix we address the stability of the results against variations in the  relative strengths of the hopping amplitudes in the 2D model. In particular, we numerically calculate the FBCs for different ratios $t_y/t_x$, see Fig. \ref{fig10}. This allows us to tune from the isotropic regime with $t_y=t_x$ to the highly anisotropic model with $t_y\ll t_x$[\onlinecite{JK1,JK6,Kane1,Kane2,Horsdal,Lederer,Gorkov,Yaro,Pawel,JK7,Oreg,Sela,Sagi,Meng}].
 The obtained slopes in the FBCs are always quantized and independent of the ratio $t_y/t_x$, and, moreover,
they are stable against disorder as long as the band gap is well-defined. Our results clearly show that the features of the FBCs as well as the resulting quantized values of the  Hall conductance are independent of the anisotropy of the model. 

\begin{figure}[!bt]
\includegraphics[width=0.75\columnwidth]{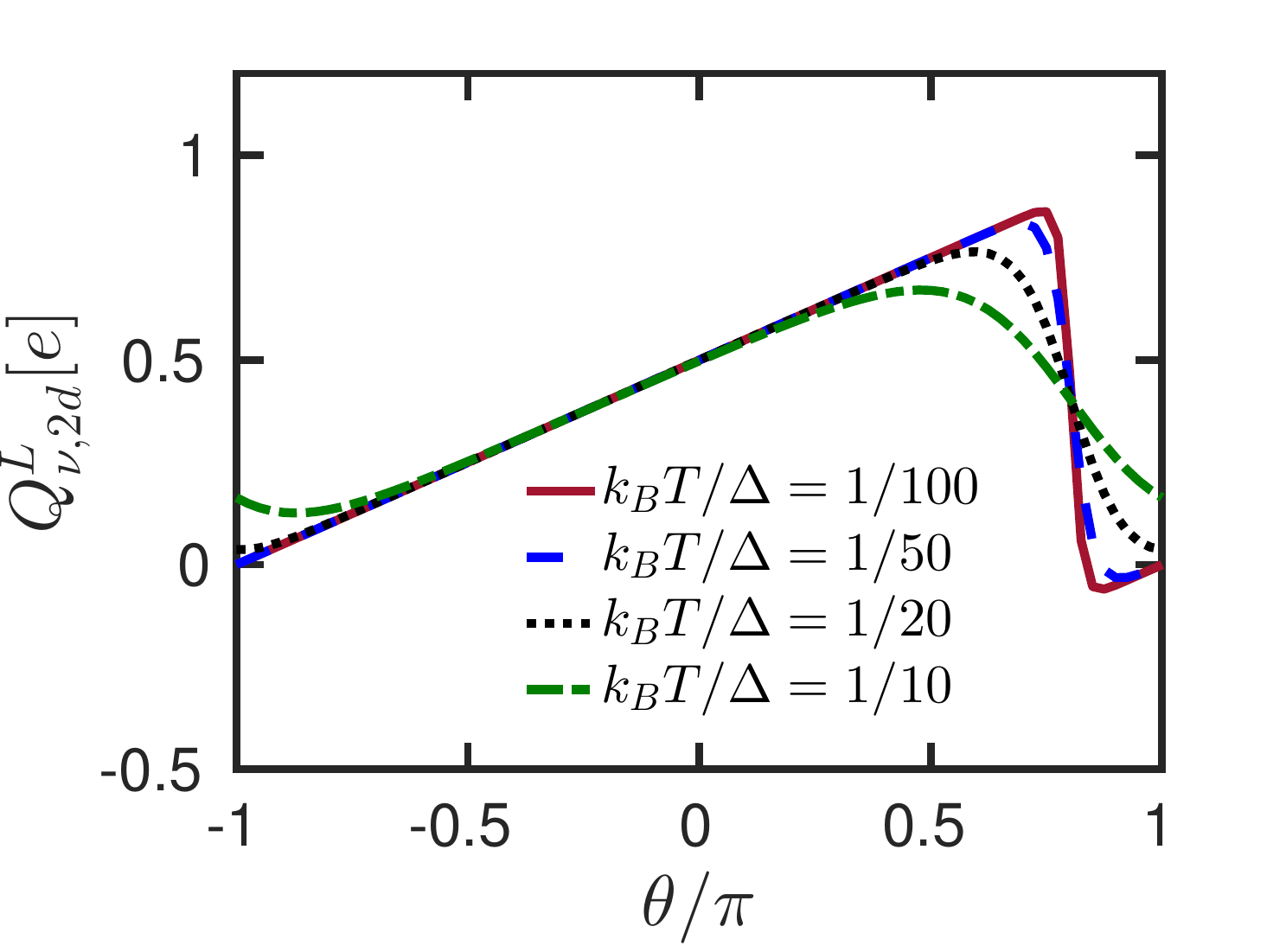}
\caption{The FBCs, $Q_{\nu, 2d}^{L}$, as a function of the flux phase $\theta$ at different temperatures: $k_BT/\Delta=1/100$ (red), 1/50 (blue), 1/20 (black), and 1/10 (green). All parameters are the same as in Fig. \ref{fig07}d. Here, $\Delta$ is the energy distance between the chemical potential and the nearest bulk band. For the chemical potential tuned inside the first ($\nu=1$) bulk gap (yellow solid line in Fig. \ref{fig02}), the energy distance to the second band is $\Delta/t=0.16$. We still observe the linear dependence of $Q_{\nu, 2d}^{L}$  on $\theta$, however, the jump in the FBC gets smoother as the temperature increases, which makes it more difficult to define the slope $c_\nu$. If $k_BT$ gets close to $\Delta$, the linear dependence of the FBC on the flux disappears.
}
\label{fig12}
\end{figure}
\section{Particle densities for $\nu=2,3$} 
\label{ap2}
In addition, we explore the profile of the local particle density $\gamma_n^\theta$ for different numbers of filled bands, see Fig.~{\ref{fig09}}. In the bulk, $\gamma_n^\theta$ does not depend on $\theta$. In contrast to that, at the boundaries, $\gamma_n^\theta$ is sensitive to the flux, giving rise to the linear slope in the $\theta$-dependence of the FBCs.

\section{FBCs at finite temperatures} 
\label{ap3}
In this Appendix, we study numerically the FBCs at finite temperature $T$, see Fig. \ref{fig12}. For this we modify the definition of the FBCs introduced in Eq. (\ref{Eq11}), where only states with negative energies contributed to the FBCs. At finite temperatures, the weight of each state with energy $\epsilon_{p}$  is given by the Fermi-Dirac distribution function $\tilde{n}(\epsilon_{p})=1/[1+\exp(\epsilon_{p}/k_B T)]$, where the index $p$ labels the $M\times N$ states of the tight-binding Hamiltonian $H^\theta$ given in Eq. (\ref{Eq8}), and $k_B$ is the Boltzmann constant. The corresponding wavefunctions are given by $\tilde \phi_p (n,m)$. The important parameters describing the effect of temperature is the energy distance $\Delta$ between the chemical potential and the nearest bulk band. If $k_B T/\Delta \ll 1$, the slope $c_\nu$ is perfectly linear. As temperature is increased, the jump in $Q_{\nu, 2d}^{s}$ gets smoother and the dependence of FBCs on the flux phase $\theta$ is linear only sufficiently far away from the jump, see Fig. \ref{fig12}. As $k_B T$ is increased further and gets close to $\Delta$, the electron 
gets thermally excited from (into) a localized edge state   into (from) extended bulk states separated by the gap $\Delta$. As a result, the definition of the FBCs assumed to be a property of the boundaries breaks down. This has an effect on $c_\nu$ as well as the quantized Hall conductance, see Fig. \ref{fig11}. The thermal broadening of the plateaus does not allow one to observe the quantization any longer if $k_B T \approx \Delta$. Thus, the plateaus will first shrink and eventually disappear as one increases the temperature.

\end{document}